\documentclass[aps,prd,floatfix,twocolumn,superscriptaddress,preprintnumbers,nofootinbib,10pt]{revtex4-1}

\pdfoutput=1
\usepackage{graphicx}
\usepackage{amssymb}
\usepackage{amsmath,amssymb}
\usepackage{color}

\usepackage{slashed}

\graphicspath{{./}{figures/}}

\definecolor{darkblue}{rgb}{0,0,0.5}
\usepackage[colorlinks,linkcolor=darkblue,citecolor=darkblue,urlcolor=darkblue]{hyperref}

\allowdisplaybreaks

\newcommand{\eqnref}[1]{Eq.~(\ref{eqn:#1})}

\newcommand{\secref}[1]{Sec.~\ref{sec:#1}}

\newcommand{\subsecref}[1]{Subsec.~\ref{subsec:#1}}

\newcommand{\appref}[1]{Appendix~\ref{sec:#1}}
\newcommand{\figref}[1]{Fig.~\ref{fig:#1}}

\begin{document}
\preprint{FERMILAB-PUB-12-034-T}


\title{New Physics Models of Direct CP Violation in Charm Decays }

\author{Wolfgang Altmannshofer}
\email{waltmann@fnal.gov}
\affiliation{Fermi National Accelerator Laboratory, P.O.~Box 500,
  Batavia, IL 60510, USA}

\author{Reinard Primulando}
\email{rprimulando@email.wm.edu}
\affiliation{Fermi National Accelerator Laboratory, P.O.~Box 500,
  Batavia, IL 60510, USA}
\affiliation{Department of Physics, College of William and Mary,
  Williamsburg, VA 23187, USA}

\author{Chiu-Tien Yu}
\email{cyu27@wisc.edu}
\affiliation{Fermi National Accelerator Laboratory, P.O.~Box 500,
  Batavia, IL 60510, USA} 
\affiliation{Department of Physics, University of Wisconsin, Madison,
  WI 53706, USA}

\author{Felix Yu}
\email{felixyu@fnal.gov}
\affiliation{Fermi National Accelerator Laboratory, P.O.~Box 500,
  Batavia, IL 60510, USA}

\begin{abstract}
In view of the recent LHCb measurement of $\Delta A_{\rm CP}$, the
difference between the time-integrated CP asymmetries in $D \to
K^+K^-$ and $D \to \pi^+\pi^-$ decays, we perform a comparative study
of the possible impact of New Physics degrees of freedom on the direct
CP asymmetries in singly Cabibbo suppressed $D$ meson decays.  We
systematically discuss scenarios with a minimal set of new degrees of
freedom that have renormalizable couplings to the SM particles and
that are heavy enough such that their effects on the $D$ meson decays
can be described by local operators. We take into account both
constraints from low energy flavor observables, in particular $D^0 -
\bar D^0$ mixing, and from direct searches.  While models that explain
the large measured value for $\Delta A_{\rm CP}$ with chirally
enhanced chromomagnetic penguins are least constrained, we identify
a few viable models that contribute to the $D$ meson decays at tree
level or through loop induced QCD penguins.  We emphasize that such
models motivate direct searches at the LHC.
\end{abstract}

\maketitle

\section{Introduction} \label{sec:intro}
%
Recently, the LHCb collaboration presented the first evidence for CP
violation (CPV) in charm quark decays~\cite{Aaij:2011in}. In
particular, a difference between the time-integrated CP asymmetries in
$D \to K^+ K^-$ and $D \to \pi^+\pi^-$
\begin{eqnarray}
\Delta A_\text{CP, LHCb} &=& A_{\rm CP}(K^+K^-) - 
A_{\rm CP}(\pi^+\pi^-) \nonumber \\
&=& (-0.82 \pm0.21 \pm0.11)\%
\end{eqnarray}
has been reported, which is non-zero at $3.5 \sigma$.  This
measurement is consistent at about the $1 \sigma$ level with the
previous measurement from CDF~\cite{Aaltonen:2011se}, and the previous
world average from the Heavy Flavor Averaging
Group~\cite{Asner:2010qj}.  The new world average, combining the LHCb
result with previous measurements of $A_{\rm CP}(K^+K^-)$ and $A_{\rm
  CP}(\pi^+\pi^-)$ at BaBar~\cite{Aubert:2007if}, Belle~\cite{:2008rx}
and CDF~\cite{Aaltonen:2011se}, is~\cite{Asner:2010qj}
\begin{equation} \label{eqn:ACP_WA}
\Delta A_\text{CP, World Average} = (-0.645 \pm 0.180)\% ~.
\end{equation}

The interpretation of this measurement as a sign of New Physics (NP)
requires a well-understood Standard Model (SM) calculation of this
observable.  Simple arguments dictate that the SM contribution to
direct CPV in $D^0$ decays must be both CKM suppressed and loop
suppressed.  Concretely, the tree level decays $D \rightarrow K^+ K^-$
and $D \rightarrow \pi^+ \pi^-$ (we implicitly include both $D^0$ and
$\bar D^0$ when discussing neutral $D$ decay modes) only involve the
first two quark generations, which cannot access the CP violating
Kobayashi-Maskawa (KM) phase.  The KM phase does enter into the
loop-induced gluon penguin diagram for singly-Cabibbo suppressed
$D^0$ decays that thus can provide both the required weak and strong
phase difference relative to the leading SM tree amplitude. This
implies that the SM prediction is loop suppressed as well as CKM
suppressed, and the na\"{i}ve expectation for direct CPV in
singly-Cabibbo suppressed $D^0$ decays is parametrically given as
$\mathcal{O}((\alpha_s / \pi)(V_{ub} V_{cb}^*)/(V_{us}
V_{cs}^*))\sim10^{-4}$. This leads to the conclusion that the LHCb
evidence of CPV at about the percent level is a sign of New Physics.

A precise SM calculation, however, is difficult to accomplish.
Although tree level and loop level SM contributions to the quark level
processes $c \rightarrow u s \bar s$ and $c \rightarrow u d \bar d$
are readily calculated, the evaluation of the hadronic matrix elements
$\langle K^+ K^- | (\bar u \Gamma_1 s) (\bar s \Gamma_2 c) | D^0
\rangle$, for example, is not easily performed.  In the simplest
approach, na\"{i}ve factorization, the hadronic matrix elements are
``factorized'' into $\langle K^+ | (\bar u \Gamma_1 s | 0 \rangle
\langle K^- | (\bar s \Gamma_2 c) | D^0 \rangle$ which is formally the
leading term in the heavy charm quark limit.  As the charm mass is
close to $\Lambda_{\text{QCD}}$, however, this approach suffers from
large $1 / m_c$ power corrections.  In particular, so-called
annihilation diagrams are ignored, where quarks are pair-produced from
the vacuum to complete the $K$ or $\pi$ mesons, as are long-range QCD
effects such as final state rescattering, where constituent $s$ quarks
of a $D \rightarrow K^+ K^-$ decay rescatter into $d$ quarks of a
$\pi^+ \pi^-$ final state.  Alternative techniques such as the
topological diagram approach organize decay and annihilation
amplitudes according to weak current insertions and $SU(3)_F$ light
quark flavor symmetry, and then try to extract amplitudes and phases
directly from $D^0$ branching ratio data.

Several recent papers have dicussed improved estimates for $\Delta
A_{\rm CP}$ in the SM.  In~\cite{Brod:2011re}, a NLO QCD factorization
calculation is amended by an estimate of the effect of certain $1/m_c$
suppressed penguin amplitudes using $D^0$ branching ratio
data. Assuming an $\mathcal{O}(1)$ strong phase, the authors find the
SM can potentially give $|\Delta A_{\rm CP}| \sim 0.4\% $.  As their
result admittedly neglects several effects which could alternatively
reduce or enhance this estimate, they conclude that the measured value
of $\Delta A_{\rm CP}$ could be reproduced in the SM.  Studies that
inform magnitudes and phases of $D^0$ meson decay amplitudes directly
from data were performed in~\cite{Pirtskhalava:2011va,Cheng:2012wr,
  Bhattacharya:2012ah} following a topological diagram approach. The
authors of~\cite{Cheng:2012wr} arrive at a slightly smaller $\Delta
A_{\rm CP} \sim -0.25\% $ estimate, which remains, they highlight,
more than 2$\sigma$ away from the world
average. In~\cite{Pirtskhalava:2011va, Bhattacharya:2012ah}, the
correlation between direct CPV in $D \rightarrow K^+ K^-$ and $\pi^+
\pi^-$ and other $D$ meson decays is emphasized as an important
cross-check of the LHCb result.

Even though there is large uncertainty in the SM value of $\Delta
A_{\rm CP}$, it is nevertheless important and exciting to consider the
possibility that we are seeing evidence of NP.  Literature prior to
the LHCb result emphasized the continued fact that CPV in the charm
sector is considered an excellent probe of NP beyond the
SM~\cite{Blaylock:1995ay, Bianco:2003vb, Nakamura:2010zzi}.  Among the
most promising probes of CPV in the charm sector are observables in
$D^0 - \bar D^0$ mixing~\cite{Golowich:2007ka, Gedalia:2009kh,
  Bigi:2009df, Kagan:2009gb, Altmannshofer:2010ad} and singly Cabibbo
suppressed $D$ decays~\cite{Grossman:2006jg, Bigi:2011re,
  Bigi:2011em}.  In fact, since flavor physics observables can probe
energy scales much higher than those directly measured, we could
potentially expect that NP at the LHC would first be seen from its
flavor effects at low energies and only later accessed directly.

Our goal is to investigate the possibility that NP is indeed
responsible for the large $\Delta A_{\rm CP}$ measurement and to
outline the corresponding NP parameter space consistent with all
experimental constraints for a variety of NP models. Some recent work
has discussed the NP possibility both model
independently~\cite{Isidori:2011qw} and in the context of various
concrete NP scenarios, including up-type flavor changing neutral
currents (FCNCs), fourth-generation fermions, $R$-parity violating
supersymmetry, and the MSSM with nonstandard sources of flavor
violation~\cite{Wang:2011uu,Rozanov:2011gj,
  Hochberg:2011ru,Chang:2012gn,Giudice:2012qq}.

Our work differs from these previous analyses since we consider a much
broader range of new NP possibilities and apply a full gamut of
experimental constraints, both from low energy experiments and
collider searches.  In particular, we systematically discuss models
with a minimal set of new degrees of freedom with renormalizable
couplings to the SM particles and are heavy enough such that their
effects on the $D$ meson decays can be described by local
operators. Specifically, we consider models with new massive neutral
gauge bosons that have flavor changing tree level couplings to quarks,
models with extended scalar sectors, and models where the $D \to K^+
K^-$ and $D \to \pi^+ \pi^-$ decays are modified at the loop level by
gluon penguins.  As discussed
in~\cite{Grossman:2006jg,Isidori:2011qw,Giudice:2012qq}, the loop
induced $\Delta F = 1$ chromomagnetic dipole operator (here and
throughout, $F$ refers to charm number except where noted) is, on
general grounds, expected to be the least constrained approach for
generating large nonstandard effects in $D^0$ meson decays. On the
other hand, the effects of four fermion operators that are, for
example, induced by tree level exchange of flavor changing NP degrees
of freedom, are highly constrained by $D^0 - \bar D^0$ mixing data. As
is well known, the $D^0 - \bar D^0$ constraints become more effective
with heavier NP degrees of freedom~\cite{Grossman:2006jg}, leading to
the expectation that almost no NP parameter space remains in models
where four fermion operators are responsible for nonstandard direct
CPV in $D \to K^+ K^-$ and $D \to \pi^+ \pi^-$.  Our work shows to
what extent this na\"{i}ve expectation holds true and identifies a few
exceptions.  We emphasize that each of the viable NP possibilities
that can accomodate a large $\Delta A_{\rm CP}$ motivates a further
phenomenological study focusing on the allowed parameter space
identified in this work, which we leave for a future study.

In~\secref{ACP}, we review aspects of CPV in neutral $D$ meson decays
that are most relevant for our analysis. In~\secref{NP}, we present
the $\Delta F = 1$ and $\Delta F = 2$ effective Hamiltonians that can
describe NP contributions to the $D \rightarrow K^+ K^-$ and $D
\rightarrow \pi^+ \pi^-$ decays and to $D^0 - \bar D^0$ mixing,
respectively. The various NP models that contribute to CPV at tree
level are discussed in~\secref{models}, while the NP models that
contribute at loop level are discussed in~\secref{penguin}. We
conclude in~\secref{conclusions}. Technical details about hadronic
matrix elements and renormalization group running, as well as a
collection of loop functions can be found in the appendices.

\section{CP Asymmetries in Neutral D Meson Decays} \label{sec:ACP}
%
The neutral $D$ meson mass eigenstates $D_1$ and $D_2$ are linear
combinations of the strong interaction eigenstates, $D^0$ and $\bar
D^0$
\begin{equation}
| D_{1,2} \rangle = p | D^0 \rangle \pm q | \bar D^0 \rangle ~.
\end{equation}
The factors $q$ and $p$ are given by
\begin{equation}
\frac{q}{p} = \sqrt{\frac{M_{12}^* - \frac{i}{2} \Gamma_{12}^*}{M_{12}
    - \frac{i}{2} \Gamma_{12}}} ~,
\end{equation}
where $M_{12}$ and $\Gamma_{12}$ are the dispersive and absorptive
part of the $D$ meson mixing amplitude. CP violation in $D$ meson
mixing is signaled by $|q/p| \neq 1$ or $\phi = \text{Arg}(q/p)
\neq 0$.

The normalized mass and width differences, $x$ and $y$, in the neutral
$D$ meson system are given by
\begin{equation}
x = \frac{\Delta M_D}{\Gamma} = 2 \tau \textnormal{Re}\left[
  \frac{q}{p} \left( M_{12} - \frac{i}{2} \Gamma_{12} \right) \right]
~,
\end{equation}
\begin{equation}
y = \frac{\Delta \Gamma_D}{2 \Gamma} = - 2 \tau \textnormal{Im}\left[
  \frac{q}{p} \left( M_{12} - \frac{i}{2} \Gamma_{12} \right) \right]
~,
\end{equation}
where the lifetime of the $D^0$ mesons $\tau = 1/\Gamma = 0.41$
ps~\cite{Nakamura:2010zzi}.

The time integrated CP asymmetry in the decay of neutral $D$ mesons to
a final CP eigenstate $f = K^+ K^-$, $\pi^+ \pi^-$ is defined as
\begin{eqnarray}
A_{\rm CP}(f) &=& \frac{\Gamma(D^0 \to f) - \Gamma(\bar D^0 \to
  f)}{\Gamma(D^0 \to f) + \Gamma(\bar D^0 \to f)} \nonumber \\ 
&=& A^m + A^i + A_f^d ~.
\end{eqnarray}
The time integrated CP asymmetry receives contributions from CPV in
mixing $A^m$, CPV in interference of decays with and without mixing
$A^i$, and from CPV in the decay itself $A_f^d$.  The ``indirect'' CP
asymmetries $A^m$ and $A^i$ are approximately independent of the final
state and depend only on $D^0 - \bar D^0$ mixing parameters
\begin{eqnarray}
A^m &=& \eta_{\rm CP}^f \frac{y}{2} \left( \left|\frac{p}{q}\right| -
\left|\frac{q}{p}\right| \right) \cos\phi ~, \\
A^i &=& \eta_{\rm CP}^f \frac{x}{2} \left( \left|\frac{p}{q}\right| +
\left|\frac{q}{p}\right| \right) \sin\phi ~,
\end{eqnarray}
where $\eta_{\rm CP}^f$ is the CP parity of the final state.  The
``direct'' CP asymmetry $A_f^d$ is instead sensitive to the final
state. The decay amplitudes of singly Cabibbo suppressed $D$ meson
decays $A(D^0 \to f) = A_f$ and $A(\bar D^0 \to f) = \bar A_f$ can be
written as~\cite{Grossman:2006jg}
\begin{eqnarray}
A_f &=& A_f^T \left( 1 + r_f e^{i (\delta_f + \phi_f)} \right) ~, \\
\bar A_f &=& \eta_{\rm CP}^f A_f^T \left( 1 + r_f e^{i (\delta_f -
  \phi_f)} \right) ~,
\end{eqnarray}
where $A_f^T$ is the dominant singly Cabibbo suppressed tree level
amplitude, which can be taken real by convention, and $r_f$ is the
relative size of subleading (``penguin'') amplitudes. With respect to
the tree amplitude, the penguin amplitudes can have a relative weak
phase $\phi_f$ and a relative strong phase $\delta_f$.

Under the assumption that $r_f$ is small, one arrives at the following
expression for the direct CP asymmetry
\begin{equation} \label{eqn:ACP_direct}
A_f^d = 2 r_f \sin\delta_f \sin\phi_f~.
\end{equation}
The difference between the time-integrated CP asymmetries in $D \to
K^+ K^-$ and $D \to \pi^+\pi^-$ measured by LHCb is given
by~\cite{Aaij:2011in}
\begin{equation}
\Delta A_{\rm CP} = A_{K^+K^-}^d - A_{\pi^+\pi^-}^d + \frac{\Delta \langle
  t \rangle}{\tau} (A^m + A^i)~,
\end{equation}
where $\Delta \langle t \rangle / \tau = (9.8 \pm 0.9)\%$ is a small
difference in the average decay times of the $D^0$ mesons in the
$K^+K^-$ and $\pi^+ \pi^-$ sample~\cite{Aaij:2011in}. Given the
existing bounds on the indirect CP asymmetries~\cite{Asner:2010qj},
the LHCb measurement of $\Delta A_{\rm CP}$ is an excellent
approximation of the difference in the direct CP asymmetries.

As already mentioned in the Introduction, charm CPV in the SM is
strongly Cabibbo suppressed. Furthermore, in the SM, direct CP
violation in $D \to K^+K^-$ and $D \to \pi^+\pi^-$ decays comes from
the interference of the tree level contribution with a loop suppressed
penguin amplitude and correspondingly, $r_f \sim \mathcal{O}(\alpha_s
/ \pi) (V_{ub} V_{cb}^*) / (V_{us} V_{cs}^*) \sim 10^{-4}$. Even
though the weak phase of the SM penguin is large ($\gamma \sim
70^\circ$) and assuming a maximal strong phase, a na\"{i}ve SM
estimate for $\Delta A_{\rm CP}$ is therefore smaller than the global
average by at least an order of magnitude.

Sizable direct CP asymmetries in the $D \to K^+K^-$ and $D \to
\pi^+\pi^-$ decays are only possible in the SM if the relevant
hadronic matrix elements are strongly
enhanced~\cite{FERMILAB-PUB-89-048-T}. Despite several recent studies
\cite{Brod:2011re, Pirtskhalava:2011va, Cheng:2012wr}, it remains
unclear to what extent such an enhancement is present and whether the
value of $\Delta A_{\rm CP}$ measured by LHCb can be explained within
the SM.

In the following we investigate the possibility that the measured
$\Delta A_{\rm CP}$ is due to New Physics.

\section{Effective Hamiltonian Approach} \label{sec:NP}

\subsection{\boldmath \texorpdfstring{$\Delta F = 1$}{Delta(F)=1} 
Effective Hamiltonian} \label{subsec:F1EFT}
%
In the New Physics frameworks discussed below, contributions to the
singly Cabibbo suppressed $D \to K^+ K^-$ and $D \to \pi^+ \pi^-$
decays can be described by the following effective Hamiltonian
\begin{eqnarray} \label{eqn:Heff_DF1}
\mathcal{H}_{\text{eff}} &=& \Big( \sum_{p} 
\lambda_p \sum_{i=1}^2\left( C_i^{(1) p} O_i^{(1) p} + \tilde C_i^{(1)
  p} \tilde O_i^{(1) p} \right) \nonumber \\
&& + \sum_{i} \left( C_i^{(1)} O_i^{(1)} + \tilde C_i^{(1)} \tilde O_i^{(1)} 
\right) \Big) ~+ {\text{ h.c.}} ~, \nonumber\\
\end{eqnarray}
where $\lambda_p = V_{cp} V_{up}^*$, and the operators $O_i^{(1)}$ are
given by
\begin{subequations}
\begin{eqnarray} 
O_1^{(1) p} &=& (\bar u p)_{V-A} (\bar p c)_{V-A} ~, \\
O_2^{(1) p} &=& (\bar u_\alpha p_\beta)_{V-A} (\bar p_\beta c_\alpha)_{V-A}
~,  \\
O_3^{(1)} &=& (\bar u c)_{V-A} \sum_{q}(\bar q q)_{V-A} ~,  \\
O_4^{(1)} &=& (\bar u_\alpha c_\beta)_{V-A} \sum_{q}(\bar q_\beta
q_\alpha)_{V-A} ~,  \\
O_5^{(1)} &=& (\bar u c)_{V-A} \sum_{q}(\bar q q)_{V+A} ~,  \\
O_6^{(1)} &=& (\bar u_\alpha c_\beta)_{V-A} \sum_{q}(\bar q_\beta
q_\alpha)_{V+A} ~,  \\
O_7^{(1)} &=& \frac{3}{2} (\bar u c)_{V-A} \sum_{q}e_q (\bar q q)_{V+A} ~,
 \\
O_8^{(1)} &=& \frac{3}{2} (\bar u_\alpha c_\beta)_{V-A} \sum_{q}e_q (\bar
q_\beta q_\alpha)_{V+A} ~,  \\
O_9^{(1)} &=& \frac{3}{2} (\bar u c)_{V-A} \sum_{q}e_q (\bar q q)_{V-A} ~,
 \\
O_{10}^{(1)} &=& \frac{3}{2} (\bar u_\alpha c_\beta)_{V-A} \sum_{q}e_q (\bar
q_\beta q_\alpha)_{V-A} ~,  \\
O_{8g}^{(1)} &=& \frac{g_s}{8\pi^2} m_c \bar u \sigma^{\mu\nu} (1+\gamma_5)
c_\beta t^a_{\alpha \beta} G_{\mu\nu}^a ~,  \\
O_{S1}^{(1)} &=& (\bar u P_L s) (\bar s P_L c) ~,  \\
O_{S2}^{(1)} &=& (\bar u_\alpha P_L s_\beta) (\bar s_\beta P_L c_\alpha) ~, \\
O_{T1}^{(1)} &=& (\bar u \sigma_{\mu\nu} P_L s) (\bar s \sigma^{\mu\nu} 
P_L c) ~,  \\
O_{T2}^{(1)} &=& (\bar u_\alpha \sigma_{\mu\nu} P_L s_\beta) 
(\bar s_\beta \sigma^{\mu\nu} P_L c_\alpha) ~.
\end{eqnarray}
\end{subequations}
The index $q$ runs over all active quark flavors, the index $p$ runs
over all active down type quark flavors, $\alpha$ and $\beta$ are
color indices (that are implicitly summed over), $e_q$ is the electric
charge of the quark $q$, $(V\pm A)$ refers to the Dirac structures
$\gamma_\mu (1 \pm \gamma_5)$, $P_{R,L}=\frac{1}{2}(1\pm\gamma_5)$ and
$\sigma_{\mu\nu} = \frac{i}{2}(\gamma_\mu \gamma_\nu - \gamma_\nu
\gamma_\mu)$. The operators $\tilde O_i^{(1) \{p \} }$ are obtained
from $O_i^{(1) \{ p \} }$ by replacing $\gamma_5 \to - \gamma_5$.

The operators $O_{1,2}^{(1) p}$ are the so-called current-current
operators. In the SM, tree level $W$ exchange generates at the
matching scale the Wilson coefficient $C_1^{(1) p} \simeq
G_F/\sqrt{2}$.  The QCD penguin operators $O_{3,4,5,6}^{(1)}$ and the
chromomagnetic operator $O_{8g}^{(1)}$ are first generated at
$\mathcal{O}(\alpha_s)$ and proportional to $V_{ub}V_{cb}^*$. The
chromomagnetic operator is proportional to the charm quark mass but
can be chirally enhanced by $v/m_c$ from NP. The QED penguin operators
$O_{7,8,9,10}^{(1)}$ are also proportional to $V_{ub}V_{cb}^*$. They
are of $\mathcal{O}(\alpha)$ and negligible in the $D \to K^+K^-$ and
$D \to \pi^+\pi^-$ decays in the SM.  In the NP models discussed below
that have tree level contributions to the $D \to K^+K^-$ and $D \to
\pi^+\pi^-$ decay amplitudes, however, the QED penguin operators can
be relevant.  The scalar operators $O_{S1,S2}^{(1)}$ become important
in the context of the 2HDM discussed in~\subsecref{2HDM} and the
scalar octet discussed in~\subsecref{octet}. The tensor operators
$O_{T1,T2}^{(1)}$ do not contribute to $D \to K^+K^-$ and $D \to
\pi^+\pi^-$ decays in na\"{i}ve factorization.  We consider them
nonetheless, because they mix with the scalar operators under
renormalization group running.

The ratio $r_f$ that enters the expression for the direct CP
asymmetry~\eqnref{ACP_direct} can be written as a function of the
Wilson coefficients appearing in~\eqnref{Heff_DF1}. We use the results
from~\cite{Grossman:2006jg} for the hadronic matrix elements that are
obtained using na\"{i}ve factorization for $O_{1,\ldots,6}^{(1)}$ and
QCD factorization~\cite{Beneke:1999br, Beneke:2000ry} for
$O_{8g}^{(1)}$.  The matrix elements obtained in na\"{i}ve
factorization are formally the leading terms in an expansion in
$\alpha_s$ and $\Lambda_{\rm QCD}/m_c$~\cite{Beneke:1999br,
  Beneke:2000ry}.  In the case of $D$ meson decays, however, and as
mentioned in the Introduction, it is known that power corrections, in
particular annihilation contributions, which are formally suppressed
by $1/m_c$, can be equally important~\cite{Grossman:2006jg,
  Brod:2011re, Pirtskhalava:2011va}. The na\"{i}ve factorization
results can therefore only be considered as rough estimates and in our
numerical analysis, we will allow for enhancements up to a plausible
factor of 3~\cite{Grossman:2006jg, Brod:2011re}.  For our analysis, we
extend the results for the hadronic matrix elements given
in~\cite{Grossman:2006jg} by including the QED penguin and scalar
operators (see~\appref{HME} for details).  We find
\begin{widetext}
\begin{eqnarray} \label{eqn:rf}
r_f e^{i \phi_f} &\simeq& \frac{1}{\lambda_p} \left(C_1^{(1) p} +
\frac{C^{(1) p}_2}{N_c} \right)^{-1} \left( \frac{\lambda_p (C_2^{(1)
    p})_{\rm NP}}{N_c} + C_4^{(1)} + \frac{C_3^{(1)}}{N_c} -
\frac{C_{10}^{(1)}}{2} - \frac{C_9^{(1)}}{2 N_c} -
\frac{3\alpha_s}{4\pi} \frac{N_c^2-1}{N_c^2} C_{8g}^{(1)}
\right. \nonumber \\
&& + \chi_f \left( C_6^{(1)} + \frac{C_5^{(1)}}{N_c} -
\frac{C_8^{(1)}}{2} - \frac{C_7^{(1)}}{2 N_c} - \frac{C_{S1}^{(1)}}{8}
 - \frac{C_{S2}^{(1)}}{8 N_c} - \frac{\alpha_s}{4\pi}
\frac{N_c^2-1}{N_c^2} C_{8g}^{(1)} \right) + ( C_i^{(1)} \leftrightarrow 
\tilde C_i^{(1)}) \Bigg)~,
\end{eqnarray}
\end{widetext}
where $N_c = 3$ is the number of colors and $p=s$, $f = K^+ K^-$ for
the $D \to K^+ K^-$ decay and $p = d$, $f = \pi^+ \pi^-$ for the $D
\to \pi^+ \pi^-$ decay.  The chiral factors $\chi_f$ are approximately
given by
\begin{equation} \label{eqn:chiralfactor}
\chi_{K^+ K^-} \simeq \frac{2 m_K^2}{m_c m_s} ~,~~
\chi_{\pi^+ \pi^-} \simeq \frac{2 m_\pi^2}{m_c(m_d + m_u)} ~,
\end{equation}
with all quark masses evaluated at the scale of the $D$ meson $\mu
\simeq m_D \simeq 1.8$ GeV. All the Wilson coefficients in~\eqnref{rf}
are evaluated at this scale. We use LO renormalization group running
for $C_{1,\ldots,10}^{(1)}$ and $C_{S1,S2,T1,T2}^{(1)}$ as well as for
$C_{8g}^{(1)}$ to evolve the Wilson coefficients from the high
matching scale, where NP degrees of freedom are integrated out, down
to $\mu \simeq m_D$. The corresponding anomalous dimensions are
collected in~\appref{gamma}. We do not include 2-loop mixing between
$C_{8g}^{(1)}$ and the other Wilson coefficients. In view of the large
uncertainties in the evaluation of the hadronic matrix elements, we
consider this approximate approach to be fully justified.

While there are no strong phase differences between the several
operator matrix elements in the na\"{i}ve factorization approach, they
can be generated by large power corrections or final state
interactions. Throughout this work, we will assume $\mathcal{O}(1)$
strong phase differences, following~\cite{Grossman:2006jg,
  Brod:2011re}.

\subsection{\boldmath \texorpdfstring{$\Delta F = 2$}{Delta(F)=2} 
Effective Hamiltonian} \label{subsec:F2EFT}
%
In the models discussed below, the most important flavor constraints
come often from $D^0 - \bar D^0$ and $K - \bar K$ mixing.  New Physics
contributions to meson mixing can be described by the effective
Hamiltonian
\begin{equation}
\mathcal{H}_{\rm eff} = \sum_{i=1}^5 C_i^{(2)} O_i^{(2)} +
\sum_{i=1}^3 \tilde{C}_i^{(2)} \tilde{O}_i^{(2)} ~+\text{ h.c.}~.
\end{equation}
In the case of $D^0 - \bar D^0$ mixing, the most important operators for our
analysis are given by
\begin{eqnarray}
O_1^{(2) D} &=& (\bar u_\alpha \gamma_\mu P_L c_\alpha)(\bar
u_\beta \gamma^\mu P_L c_\beta)~, \nonumber \\
\tilde{O}_1^{(2) D} &=& (\bar u_\alpha \gamma_\mu P_R c_\alpha)(\bar
u_\beta \gamma^\mu P_R c_\beta)~, \nonumber \\
\tilde{O}_2^{(2) D} &=& (\bar u_\alpha P_R c_\alpha)(\bar
u_\beta P_R c_\beta)~.
\end{eqnarray}
In the case of $K - \bar K$ mixing, the operators most relevant for our
analysis are
\begin{eqnarray}
O_1^{(2) K} & = & (\bar d_\alpha \gamma_\mu P_L
s_\alpha)(\bar d_\beta \gamma^\mu P_L s_\beta)~, \nonumber \\
\tilde{O}_1^{(2) K} & = & (\bar d_\alpha \gamma_\mu P_R
s_\alpha)(\bar d_\beta \gamma^\mu P_R s_\beta)~, \nonumber \\
O_4^{(2) K} & = &
(\bar d_\alpha P_L s_\alpha)(\bar d_\beta P_R s_\beta)~, \nonumber \\
O_5^{(2) K} & = &
(\bar d_\alpha P_L s_\beta)(\bar d_\beta P_R s_\alpha)~.
\end{eqnarray}
In the above expressions, $P_{R,L}=\frac{1}{2}(1\pm\gamma_5)$ and
$\alpha,\beta$ are color indices (that are implicitly summed over).

The Wilson coefficients $C_i^{(2)}$ are again obtained by integrating
out the NP degrees of freedom at a scale of the order of the mass of
the new particles. Using renormalization group
evolution~\cite{Ciuchini:1997bw, Buras:2000if}, these coefficients are
subsequently run down to the low scale where the hadronic matrix
elements~\cite{Becirevic:2001xt, Babich:2006bh, Bona:2007vi} are
given. Combining Wilson coefficients with the hadronic matrix elements
gives the NP contribution to the dispersive part of the mixing
amplitude $M_{12}$.\footnote{The absorptive part of the mixing
  amplitude $\Gamma_{12}$ is not sensitive to new short distance
  dynamics.} In the case of $D^0 - \bar D^0$ mixing, the SM
contributions to neither the dispersive part nor the absorptive part
of the mixing amplitude can be predicted reliably as they are
dominated by long distance effects~\cite{Falk:2001hx,Falk:2004wg}. In
our numerical analysis, we allow the long distance contributions to
vary in the ranges $M_{12}^{\rm LD} \in [-0.02 , 0.02]$ ps$^{-1}$ and
$\Gamma_{12}^{\rm LD} \in [-0.04 , 0.04]$
ps$^{-1}$~\cite{Ciuchini:2007cw}, such that by themselves they can
saturate the experimental values. We apply the most recent averages
and $1\sigma$ errors of the experimental results on the $D^0 - \bar
D^0$ mixing parameters~\cite{Asner:2010qj}
\begin{eqnarray} 
x = (0.63^{+0.19}_{-0.20})\% ~&,&~~ 
y = (0.75 \pm 0.12)\% ~, \nonumber \\ 
\left| q/p \right| = 0.89^{+0.17}_{-0.15} ~&,&~~  
\phi = (-10.1^{+9.4}_{-8.8})^\circ ~,
\end{eqnarray}
at the $2\sigma$ level throughout our analysis.

We note that in many of our minimal field content scenarios, the NP
vertices used in the $\Delta F = 1$ operators are also used for the
$\Delta F = 2$ operators, leading to a phase relation $2 \phi_{F = 1}
= \phi_{F = 2}$ between the CPV for $D^0$ decays and the CPV for $D^0
- \bar D^0$ mixing.  This relation implies that the $D^0 - \bar D^0$
mixing constraint is best satisfied by eliminating the CPV in $D^0 -
\bar D^0$ mixing and saturating the $D^0 - \bar D^0$ mixing transition
amplitude.  On the other hand, in non-minimal constructions, this
phase relation could be different, possibly making the null
observation of CPV in $D^0 - \bar D^0$ mixing the more restrictive
constraint.

\section{New Physics Contributions at Tree Level} \label{sec:models}
%
We concentrate on New Phyiscs models where the new degrees of freedom
are heavy enough such that their effects in low energy observables can
be reliably described by the local operators introduced
in~\secref{NP}.  We do not consider scenarios with very light
mediators, which is beyond the scope of this work.  Moreover, we focus
on models where the new degrees of freedom have renormalizable
couplings to SM degrees of freedom.

In this section, we analyze models where at most one new field is
added to the SM.  We first discuss extensions of the SM in which a
massive neutral gauge boson leads to tree level contributions to the
$D \to K^+ K^-$ and $D \to \pi^+ \pi^-$ decay ampitudes. We consider a
flavor changing coupling of the SM $Z$ boson in~\subsecref{Z}, a
flavor changing $Z^\prime$ in~\subsecref{Zprime} and a flavor changing
heavy gluon in~\subsecref{Gprime}. We also comment on the possible
effects of a new charged gauge boson in~\subsecref{Wprime}. Then, we
analyze models with extended scalar sectors, namely a 2 Higgs doublet
model with Minimal Flavor Violation in~\subsecref{2HDM}, a model with
a scalar octet in~\subsecref{octet} and a model with a scalar diquark
in~\subsecref{diquark}.

Models that contain more than one non-SM particle and where NP
contributions to the $D$ meson decays are first generated at the one
loop level are discussed in \secref{penguin}.

\subsection{\boldmath Flavor Changing \texorpdfstring{$Z$}{Z}} 
\label{subsec:Z}
%
\begin{figure}[tbp]
\centering \includegraphics[width=0.22\textwidth]{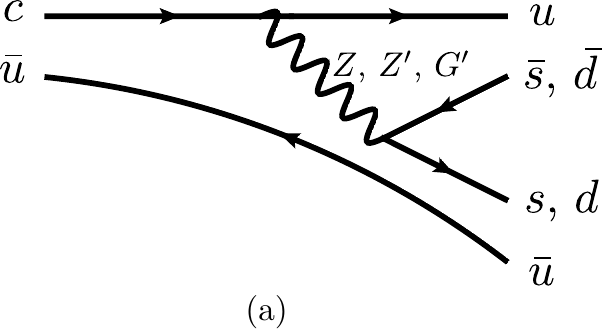} ~~~~
\includegraphics[width=0.22\textwidth]{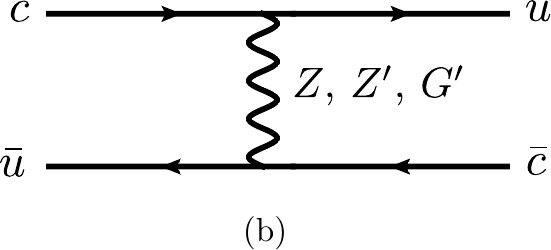}
\caption{Example tree level Feynman diagrams that contribute~(a) to
  the $D \to K^+K^-$ and $D \to \pi^+\pi^-$ decay amplitudes and~(b)
  to $D^0 - \bar D^0$ mixing in the cases of a flavor changing $Z$,
  $Z^\prime$ and heavy gluon $G^\prime$.}
\label{fig:Diagrams_ZZpGp}
\end{figure}
%
We consider a flavor changing coupling of the SM $Z$ boson to the
right-handed charm and up quark
\begin{equation}
\mathcal{L}_{\text{int}} = X_{cu} \bar c_R \gamma^\mu u_R Z_\mu ~+
\text{ h.c.}~,
\end{equation}
where $X_{cu}$ is a complex parameter. A complementary setup, where
flavor changing couplings involving the top quark generate an
effective $c \to u$ transition at the loop level by a double flavor
flip $c \to t \to u$, is discussed in~\cite{Giudice:2012qq}. Flavor
changing $Z$ couplings can appear in various
scenarios~\cite{Buchalla:2000sk}, for example in models with
non-sequential generations of quarks~\cite{Nir:1990yq} and also in
models with extra $U(1)$ gauge symmetries~\cite{Langacker:2000ju}, or
can be loop induced as in SUSY models. In the absence of $SU(2)_L$
breaking sources, the $\bar c_R u_R Z$ coupling has the form of a
charge radius interaction and vanishes for $q^2 \to 0$, where $q$ is
the momentum of the $Z$ boson. The dominant contribution to the
coupling $X_{cu}$ is therefore in general expected to be proportional
to $v^2/\Lambda_{\text{NP}}^2$, where $v$ is the Higgs vacuum
expectation value (vev) and $\Lambda_{\text{NP}}$ is the NP scale
where the flavor changing $Z$ coupling is generated.

As shown in diagram~(a) of~\figref{Diagrams_ZZpGp}, the $X_{cu}$
coupling leads to tree level contributions to the Wilson coefficients
$\tilde C_5^{(1)}$, $\tilde C_7^{(1)}$ and $\tilde C_9^{(1)}$
\begin{eqnarray}
\tilde C_5^{(1)} &=& - \frac{1}{3} \frac{g}{2c_W}
\frac{X_{cu}^*}{4 M_Z^2} ~,\nonumber \\
\tilde C_7^{(1)} &=& \frac{2}{3} g c_W 
\frac{X_{cu}^*}{4 M_Z^2} ~,\nonumber \\
\tilde C_9^{(1)} &=& - \frac{2}{3} \frac{g s_W^2}{c_W}
\frac{X_{cu}^*}{4 M_Z^2} ~.
\end{eqnarray}
The flavor changing $\bar c u Z$ coupling also inevitably generates
tree level contributions to $D^0 - \bar D^0$ mixing
\begin{equation}
\tilde{C}_1^{(2) D} = \frac{(X_{cu}^*)^2}{2 M_Z^2} ~. 
\end{equation}
If the $Z$ boson has flavor changing couplings to left-handed quarks,
 1-loop contributions to $\epsilon^\prime/\epsilon$ would also be
generated. In order to avoid this constraint, we restrict ourselves to the
$\bar c_R u_R Z$ coupling.

\begin{figure}[tbp]
\centering
\includegraphics[width=0.45\textwidth]{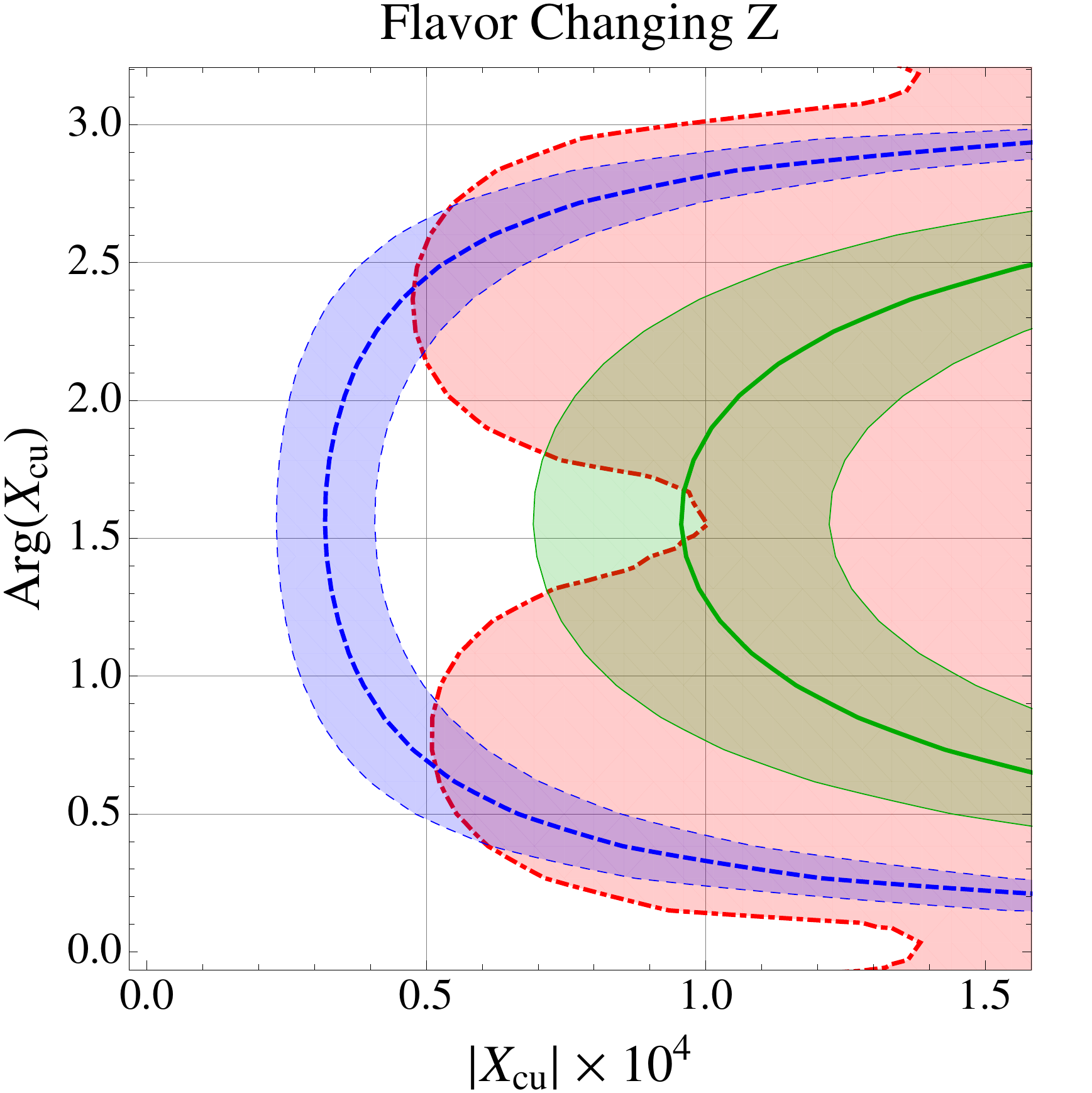}
\caption{Regions in the $|X_{cu}|$ -- Arg$(X_{cu})$ plane compatible
  with the data on $\Delta A_\text{CP, WA}$ at the $1\sigma$ level in the
  model with a flavor changing $Z$. The green (solid) band corresponds
  to the expressions for the decay amplitude in na\"{i}ve
  factorization, the blue (dashed) band assumes an enhancement of the
  hadronic matrix elements by a factor of 3.  The red (dash-dotted)
  region is excluded by the $D^0 - \bar D^0$ mixing constraints.}
\label{fig:Z}
\end{figure}

In~\figref{Z}, we show the regions in the $|X_{cu}|$ -- Arg$(X_{cu})$
plane that are compatible with the range for $\Delta A_{\rm CP}$ in
\eqnref{ACP_WA} at the $1\sigma$ level. The green (solid) band is
obtained using the expressions for the decay amplitude in na\"{i}ve
factorization. The blue (dashed) band assumes an enhancement of the
hadronic matrix elements by a factor of 3. The red (dash-dotted)
region is excluded by the constraints from $D^0 - \bar D^0$
mixing. The $D^0 - \bar D^0$ constraints are minimized for
Arg$(X_{cu}) = 0$, $\pi/2$, $\pi$, $3\pi/2$, where constraints from
CPV in $D^0 - \bar D^0$ are not effective and the dominant constraint
comes from the normalized mass difference $x$. Indeed, if no
enhancement of the hadronic matrix elements is assumed, sizeable NP
effects in $\Delta A_{\rm CP}$ are only compatible with $D^0 - \bar
D^0$ mixing in a small corner of parameter space with Arg$(X_{cu})
\simeq \pi/2, 3\pi/2$.  Still, barring the finetuned situations
Arg$(X_{cu}) = \pi/2, 3\pi/2$, sizeable NP effects in $\Delta A_{\rm
CP}$ also imply indirect CPV in $D^0 - \bar D^0$ mixing close to the
current experimental bounds.  The required size of the flavor changing
coupling $v^2/\Lambda_{\rm NP}^2 \propto |X_{cu}| \simeq 10^{-4}$
points towards a NP scale of $\Lambda_{\rm NP} \lesssim \text{few}
\times 10$~TeV, where this coupling is generated and not necessarily within the
immediate reach of direct searches.

\subsection{\boldmath Flavor Changing \texorpdfstring{$Z^\prime$}{Z'}} 
\label{subsec:Zprime}
%
\begin{figure*}[tbp]
\centering \includegraphics[width=0.45\textwidth]{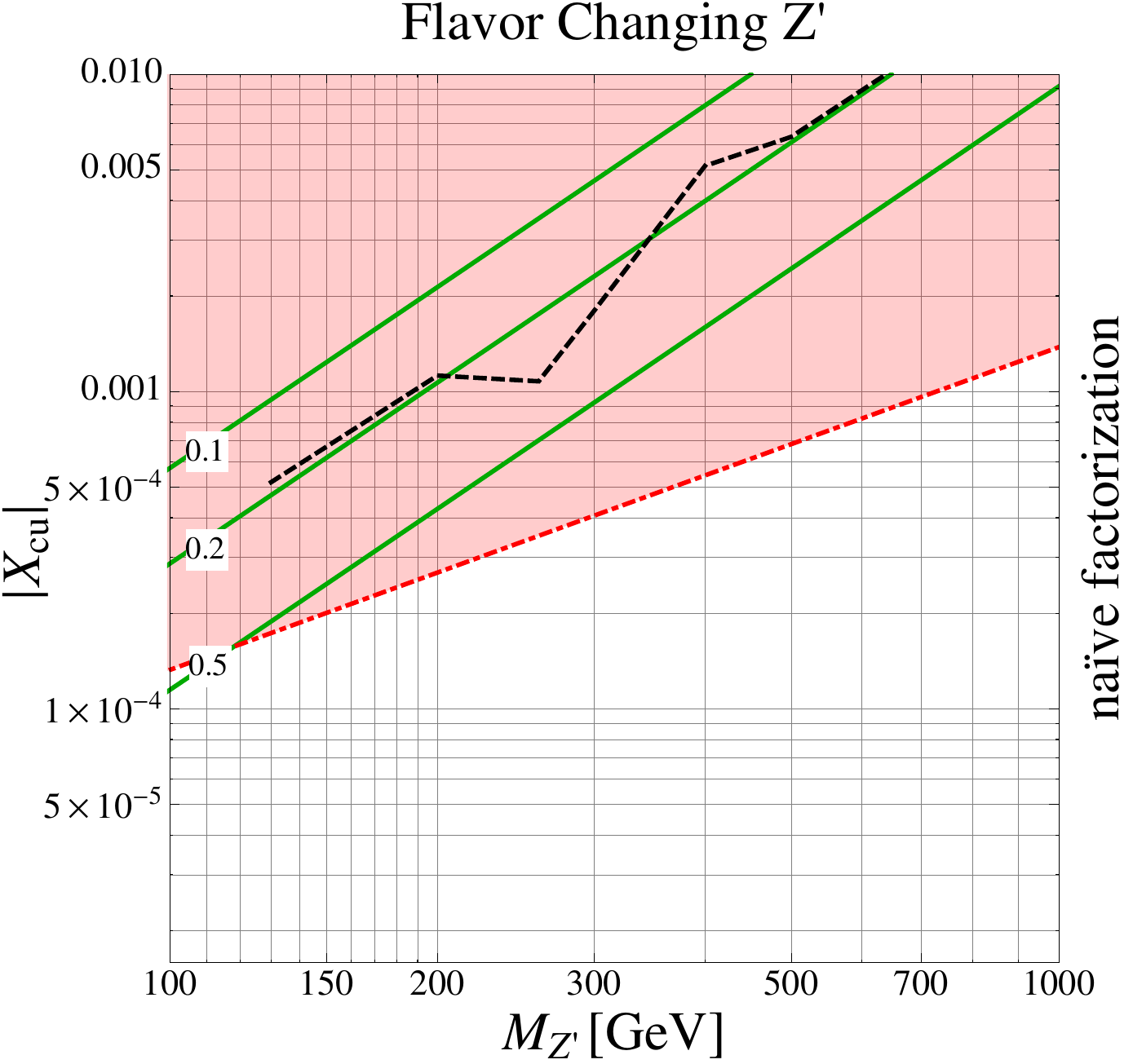} ~~~
\includegraphics[width=0.45\textwidth]{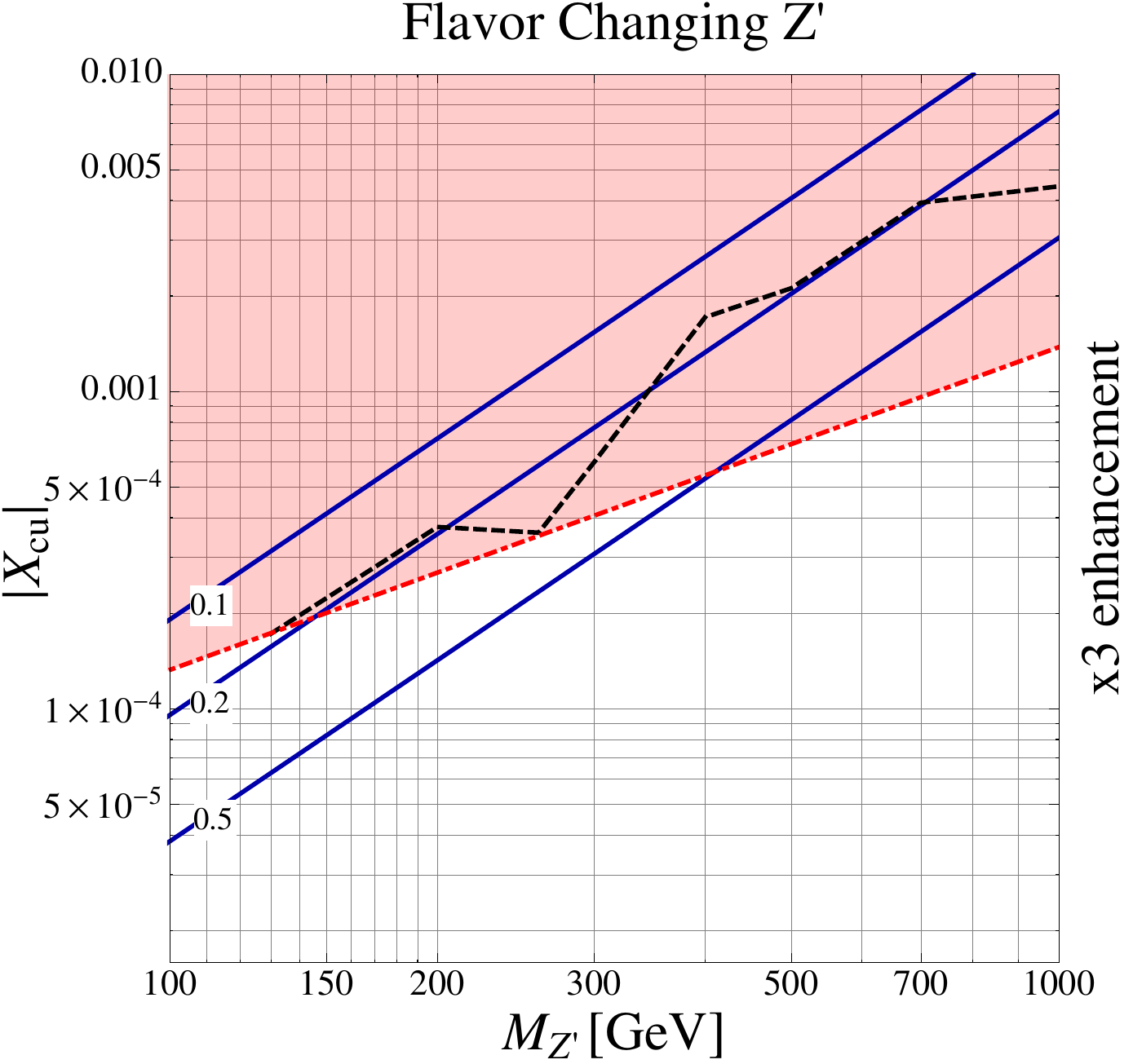}
\caption{The $M_{Z^\prime} - |X_{cu}|$ plane, setting Arg$(X_{cu}) =
  \pi/2$.  In the left plot, $\Delta A_{\rm CP}$ is evaluated in
  na\"{i}ve factorization and in the right plot, we allow for an
  enhancement by a factor of 3.  Along the solid green and blue lines
  the NP contributions to $\Delta A_{\rm CP}$ match the world average. 
  The different green and blue lines correspond to different
  common choices of the flavor conserving couplings $g_u = g_d = g_L$
  as indicated. The black dashed line shows the constraint on the
  flavor conserving coupling from dijet searches.  The red
  (dash-dotted) region is excluded by $D^0 - \bar D^0$ mixing
  constraints.}
\label{fig:Zprime}
\end{figure*}
%
Next, we consider a leptophobic massive $Z^\prime$ gauge boson with
tree level flavor changing couplings to right-handed up and charm
quarks. Models of this type can be easily constructed if the
$Z^\prime$ couples with SM degrees of freedom through higher
dimensional operators~\cite{Fox:2011qd}. Flavor changing couplings of
a $Z^\prime$ can also arise, for example, in models with family
non-universal couplings~\cite{Langacker:2000ju, Arhrib:2006sg}. We
parameterize the interactions of the $Z^\prime$ in the following way
\begin{eqnarray}
\mathcal{L}_{\text{int}} &=& g_L \bar u_L^i \gamma^\mu u_L^i 
Z_\mu^\prime + g_u \bar u_R^i \gamma^\mu u_R^i Z_\mu^\prime 
\nonumber \\
&& + g_L \bar d_L^i \gamma^\mu d_L^i Z_\mu^\prime + g_d \bar d_R^i
\gamma^\mu d_R^i Z_\mu^\prime \nonumber \\
&& + X_{cu} \bar c_R \gamma^\mu u_R Z^\prime_\mu ~+ \text{ h.c.} ~,
\end{eqnarray}
where the flavor universal couplings $g_u$, $g_d$ and $g_L$ are free
real parameters and the (small) flavor changing coupling $X_{cu}$ is a
free, complex parameter.  We restrict ourselves to a tree level $c \to
u$ coupling and do not consider $t \to c$ and $t \to u$ couplings that
could induce the $c \to u$ transition at the loop level.

Depending on whether the flavor conserving couplings of the $Z^\prime$
are to left-handed or right-handed quarks, the flavor changing $\bar c
u Z^\prime$ coupling can induce tree level contributions to the Wilson
coefficients $\tilde C_5^{(1)}$, $\tilde C_3^{(1)}$, and $\tilde
C_9^{(1)}$
\begin{eqnarray}
\tilde C_3^{(1)} &=& \frac{(g_u + 2
  g_d)}{3}\frac{X_{cu}^*}{4 M_{Z^\prime}^2} ~,\nonumber \\
\tilde C_9^{(1)} &=& \frac{2(g_u - g_d)}{3} \frac{X_{cu}^*}{4 M_{Z^\prime}^2}
~, \nonumber \\ 
\tilde C_5^{(1)} &=& \frac{g_L X_{cu}^*}{4 M_{Z^\prime}^2}~.
\end{eqnarray}
If we assume $g_u = g_d$, then only contributions to $\tilde
C_3^{(1)}$ and $\tilde C_5^{(1)}$ are generated. The contributions
from $\tilde C_{3,5,9}^{(1)}$ to the $D \to K^+ K^-$ and $D \to \pi^+
\pi^-$ decay amplitudes are color suppressed. The contribution from
$\tilde C_5^{(1)}$ is helicity enhanced.

The flavor changing $\bar c u Z^\prime$ coupling also induces tree
level contributions to $D^0 - \bar D^0$ mixing
\begin{equation}
\tilde{C}_1^{(2) D} = \frac{(X_{cu}^*)^2}{2 M_{Z^\prime}^2} ~. 
\end{equation}
We highlight the following point: as both the NP contributions to the
$\Delta F =2$ mixing amplitude and the $\Delta F =1$ decay amplitudes
are described by dimension 6 operators, they decouple with the NP mass
squared. Yet while the $\Delta F =2$ amplitude is obviously
proportional to the square of the flavor changing coupling, the
$\Delta F =1$ amplitude is linearly proportional in this
coupling. Correspondingly, the constraint from $D^0 - \bar D^0$ mixing
becomes more effective with heavier NP mass. Analogous arguments hold
in all the other NP scenarios discussed in this work.

If the $Z^\prime$ boson has flavor changing couplings to left-handed
quarks, tree level contributions to $K - \bar K$ mixing would be
unavoidably generated since the $\bar c_L u_L Z^\prime$ and $\bar s_L
d_L Z^\prime$ couplings are related by the CKM matrix due to $SU(2)_L$
invariance. Since constraints coming from $K - \bar K$ mixing are
considerably stronger than those coming from $D^0 - \bar D^0$ mixing,
we restrict ourselves to the $\bar c_R u_R Z$ coupling. As a result,
1-loop contributions to $\epsilon^\prime/\epsilon$ that can lead to
constraints are also absent.

The dijet searches at hadron colliders set additional constraints on
the model. In this paper, we consider the searches at
UA2~\cite{Alitti:1993pn}, CDF~\cite{Abe:1997hm} and
CMS~\cite{Chatrchyan:2011ns}. The UA2 collaboration probed the light
dijet mass region, from 130 GeV to 300 GeV, while the CDF search
covers a dijet mass range from 260 GeV to 1.4 TeV. A higher mass
range, $1.0 - 4.1$ TeV, is probed by the CMS experiment. There are
other dijet searches from the D\O\ ~\cite{Abazov:2003tj} and ATLAS
experiments~\cite{Aad:2011fq}. The D\O\ collaboration analyzed $109
\textrm{ pb}^{-1}$ of data, however, while CDF analyzed $1.13 \textrm{
fb}^{-1}$, indicating the D\O\ bound is less competitive than the one
from CDF. The ATLAS bound is expected to be comparable with the CMS
bound since both of the experiments analyzed $1 \textrm{ fb}^{-1}$ of
data.

We simulate the $Z^\prime$ production using \verb|MadGraph 5|
\cite{Alwall:2011uj}, and the width of the $Z^\prime$ is calculated
with \verb|CompHEP|~\cite{Pukhov:1999gg}, varying $g_u = g_d =
g_L$. We compare the simulated cross section with the limit on dijet
production from UA2 (Fig.~2 of~\cite{Alitti:1993pn}), CDF (Table I
of~\cite{Abe:1997hm}) and CMS (Table 1 of~\cite{Chatrchyan:2011ns}).
In calculating the bound, we ignore $X_{cu}$ since it is at least one
order of magnitude smaller than $g_u$, $g_d$ and $g_L$.

The plots in~\figref{Zprime} show the $M_{Z^\prime} - |X_{cu}|$ plane,
setting Arg$(X_{cu}) = \pi/2$. In the left plot, $\Delta A_{\rm CP}$
is evaluated in the na\"{i}ve factorization approach, while in the
right plot we allow for an enhancement of the hadronic matrix elements
by a factor of 3.  Along the green (solid, left plot) and blue (solid,
right plot) lines, the NP contributions to $\Delta A_{\rm CP}$ match
the world average. The different green or blue lines correspond
to different choices of the flavor conserving couplings $g_u = g_d =
g_L = 0.1$, $0.2$, or $0.5$. The region below the black (dashed) line
requires a flavor conserving coupling that is excluded by dijet
searches.  The red (dash-dotted) region is excluded by the constraints
from $D^0 - \bar D^0$ mixing. The choice Arg$(X_{cu}) = \pi/2$
corresponds to a maximal phase for the NP contributions to the $D \to
K^+ K^-$ and $D \to \pi^+ \pi^-$ decays while simultaneously
minimizing the constraint from $D^0 - \bar D^0$ mixing. Choosing
Arg$(X_{cu}) = \pi/3$ would lead to $\mathcal{O}(1)$ phases both in
the decays and in $D^0 - \bar D^0$ mixing, and the corresponding
constraint would be more stringent by a factor of $\sim 2$. As
expected, the $D^0 - \bar D^0$ constraint becomes more effective with
larger $Z^\prime$ mass.

We observe that even allowing for an enhancement in $\Delta A_{\rm
CP}$ by a factor of three, the $D^0 - \bar D^0$ mixing constraint in
combination with dijet searches rules out a $Z^\prime$ as a tree level
NP explanation for the measured $\Delta A_{\rm CP}$.  We do not
consider $Z^\prime$ masses below 100 GeV, which would be constrained
from $Z - Z^\prime$ mixing~\cite{Langacker:2008yv}, but the exact
constraints would be model dependent and are beyond the scope of this
work.

\subsection{Flavor Changing Heavy Gluon} \label{subsec:Gprime}
%
\begin{figure*}[tb]
\centering \includegraphics[width=0.45\textwidth]{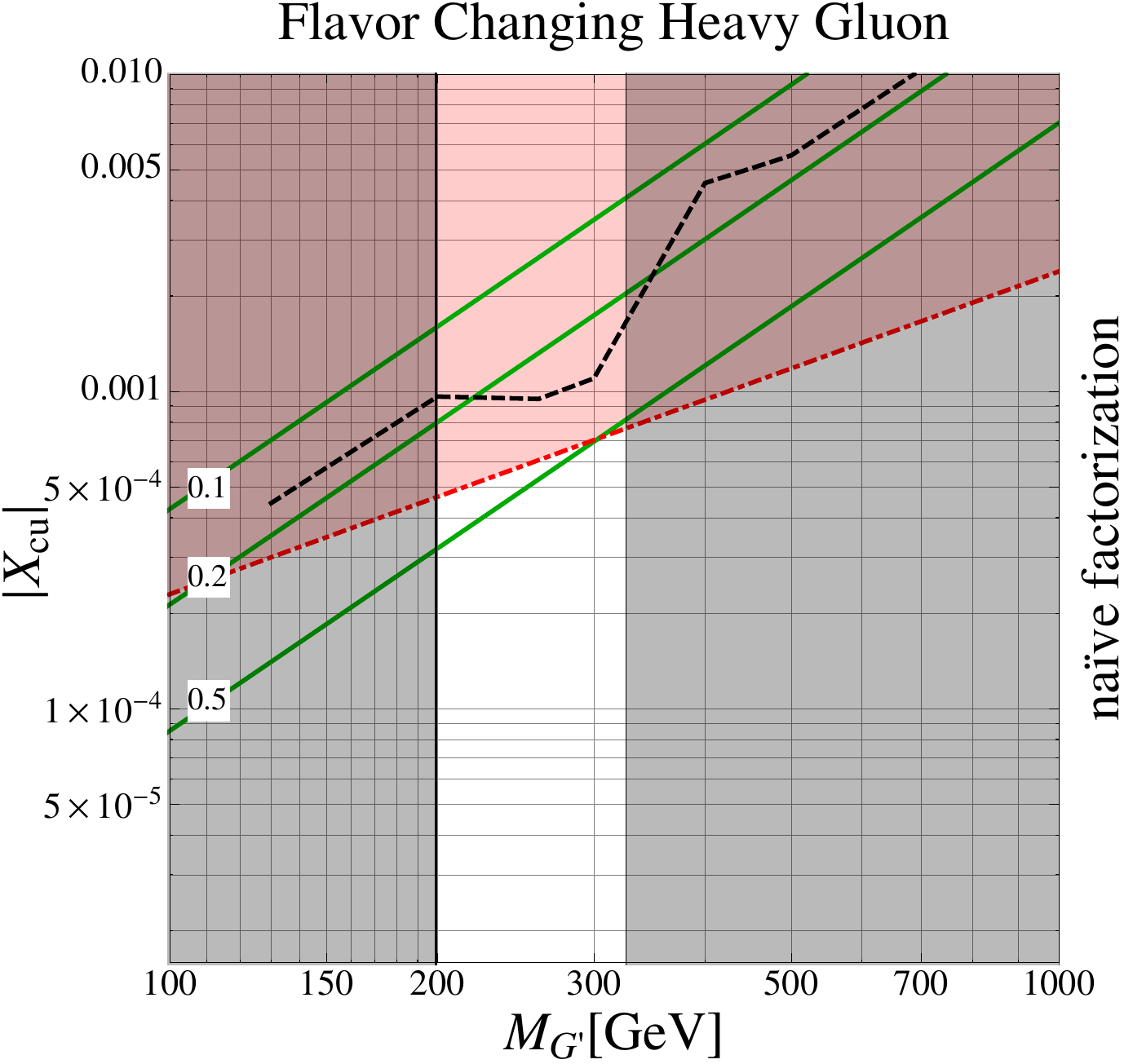} ~~~
\includegraphics[width=0.45\textwidth]{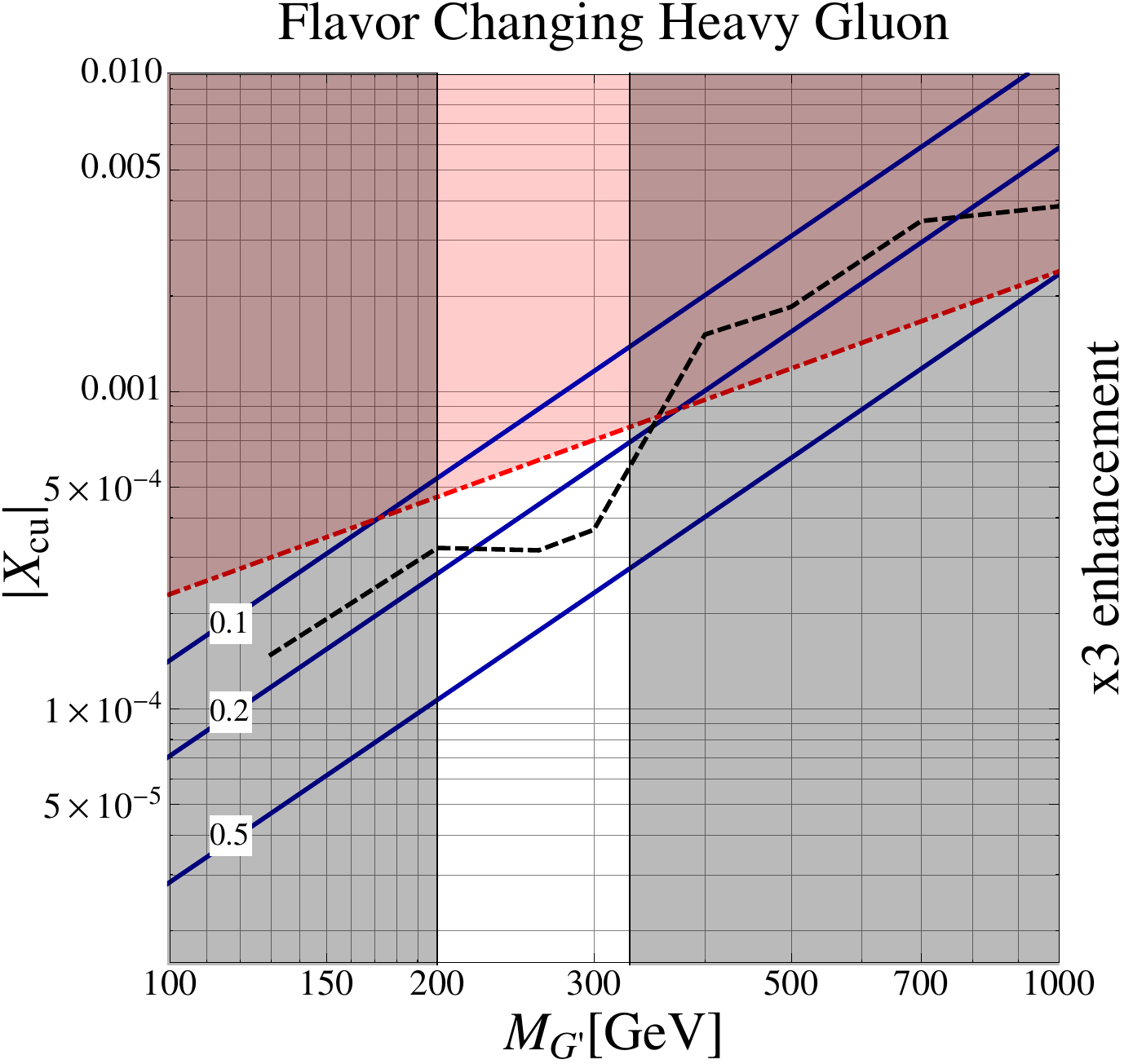}
\caption{As~\figref{Zprime} but in the model with flavor changing
heavy gluon.  The shaded vertical bands are excluded by dijet pair
searches at ATLAS (left band) and CMS (right band).}
\label{fig:Gprime}
\end{figure*}

Heavy color octet vector bosons that have flavor changing couplings to
SM quarks can arise, for example, in models with warped extra
dimensions~\cite{Agashe:2004cp} and also in models of axigluons with
family non-universal couplings~\cite{Haisch:2011up}.  We concentrate
again on a direct tree level coupling between right-handed up and
charm quarks
\begin{eqnarray}
\mathcal{L}_{\text{int}} &=& g_L \bar u_L^i \gamma^\mu T^a u_L^i 
(G^\prime)_\mu^a +
g_u \bar u_R^i \gamma^\mu T^a u_R^i (G^\prime)_\mu^a \nonumber \\
&& + g_L \bar d_L^i \gamma^\mu T^a d_L^i (G^\prime)_\mu^a + g_d \bar
d_R^i \gamma^\mu T^a d_R^i (G^\prime)_\mu^a \nonumber \\
&& + X_{cu} \bar c_R \gamma^\mu T^a u_R (G^\prime)_\mu^a ~+ \text{ h.c.} ~.
\end{eqnarray}
The flavor universal couplings $g_u$, $g_d$ and $g_L$ are free real
parameters: the (small) flavor violating coupling $X_{cu}$ is a free,
complex parameter.

The heavy gluon can generate tree level contributions to $\tilde
C_{3,4,5,6}^{(1)}$ and $\tilde C_{9,10}^{(1)}$
\begin{eqnarray}
\tilde C_4^{(1)} &=& \frac{(g_u + 2 g_d)}{3}
\frac{X_{cu}^*}{8 M_{G^\prime}^2} ~,~~ \tilde C_3^{(1)} = \frac{-1}{N_c}
\tilde C_4^{(1)} ~, \nonumber \\
\tilde C_{10}^{(1)} &=& \frac{2(g_u - g_d)}{g_u + 2g_d} \tilde C_4^{(1)} ~,~~
\tilde C_9^{(1)} = \frac{-1}{N_c} \tilde C_{10}^{(1)} ~, \nonumber \\
\tilde C_6^{(1)} &=& \frac{g_L X_{cu}^*}{8
  M_{G^\prime}^2} ~,~~ \tilde C_5^{(1)} = \frac{-1}{N_c} \tilde C_6^{(1)} ~.
\end{eqnarray}
To generate $\tilde C_{3,4}^{(1)}$ and $\tilde C_{9,10}^{(1)}$, the RH
flavor conserving coupling is required, while for $\tilde
C_{5,6}^{(1)}$ the LH flavor conserving coupling is required. If we
assume $g_u = g_d$, $\tilde C_{9,10}^{(1)}$ are absent. The
Wilson coefficients $\tilde C_{3,5,9}^{(1)}$ are color suppressed, and
furthermore, their contributions to the decay amplitudes are color
suppressed. The contributions to the decay amplitudes from $\tilde
C_5^{(1)}$ and $\tilde C_6^{(1)}$ are helicity enhanced.  Due to the
presence of $\tilde C_6^{(1)}$, we expect slightly larger NP
contributions to $\Delta A_{\rm CP}$ than the $Z^\prime$ scenario.

Similar to the models discussed before, the flavor changing $\bar c u
G^\prime$ coupling also leads to tree level contributions to $D^0 -
\bar D^0$ mixing
\begin{equation}
\tilde{C}_1^{(2) D} = \frac{1 - N_c}{2 N_c} \frac{(X_{cu}^*)^2}{2
  M_{G^\prime}^2} ~.
\end{equation}
Compared to the $Z^\prime$ case, the contribution to $D^0 - \bar D^0$
mixing is suppressed by a factor $|(1 - N_c) /(2N_c)| = 1/3$. As
discussed for the $Z^\prime$ case, we do not consider flavor changing
couplings to left-handed quarks to avoid the stringent constraints
from $K - \bar K$ mixing and $\epsilon^\prime/\epsilon$.

Collider constraints come again from dijet searches at the hadron
colliders, and we evaluate them in a similar fashion to the $Z^\prime$
case. Additional constraints arise from recent results on four jet
searches at the LHC, searching for pair production of dijet
resonances. Pair production of the heavy gluon is fixed by QCD to good
approximation and the production cross section depends only on the
$G^\prime$ mass~\cite{Dobrescu:2007yp}. The mass range from 100 GeV to
200 GeV is covered by an ATLAS search using 34
pb$^{-1}$~\cite{Aad:2011yh} while a CMS search using 2.2
fb$^{-1}$~\cite{CMS-EXO-11-016} starts at 320 GeV.  The intermediate
region from 200 GeV to 320 GeV is not considered in the CMS search
because of the multijet trigger turn-on effects on the QCD background
fit curve, which well models the QCD multijet background above 320
GeV. We simulate pair production of the $G^\prime$ resonance using
\verb|Madgraph 5|. We find that a $G^\prime$ mass in the ranges
100~GeV $< M_{G^\prime} <$ 200~GeV is excluded. Assuming the signal
acceptance to be 3\% (13\%), which corresponds to the lowest (highest)
signal efficiency found in~\cite{CMS-EXO-11-016}, we find that a
$G^\prime$ mass from 320~GeV up to 720~GeV (1000~GeV) is excluded.  We
conservatively use the stronger bound derived from an acceptance of
13\% in the following discussion.

The plots in~\figref{Gprime} show the $M_{G^\prime} - |X_{cu}|$ plane
with Arg$(X_{cu}) = \pi/2$ in order to minimize the constraint from
$D^0 - \bar D^0$ mixing. In the left plot, $\Delta A_{\rm CP}$ is
evaluated in the na\"{i}ve factorization approach, while in the right
plot we allow for an enhancement by a factor of 3. The NP
contributions to $\Delta A_{\rm CP}$ match the world average
along the green (solid, left plot) and blue (solid, right plot)
lines. The different green or blue lines correspond to different
choices of the flavor conserving couplings $g_u = g_d = g_L = 0.1$,
$0.2$, or $0.5$. The region below the black (dashed) line requires a
flavor conserving coupling that is excluded by dijet resonance
searches.  The red (dash-dotted) region is excluded by the constraints
from $D^0 - \bar D^0$ mixing.  The vertical bands are excluded by the
dijet pair searches at ATLAS (left band) and CMS (right band).

Because of different $\mathcal{O}(1)$ factors in the $D \to K^+ K^-$
and $D \to \pi^+\pi^-$ decay amplitudes and $D^0 - \bar D^0$ mixing
compared to the $Z^\prime$, a heavy gluon appears slightly better
suited to generate nonstandard effects in $\Delta A_{\rm CP}$.  Yet
only after allowing for an enhancement in $\Delta A_{\rm CP}$ by a
factor of 3 can the combined constraints from $D^0 - \bar D^0$ mixing
and dijet searches be made compatible with the measured $\Delta A_{\rm
CP}$.  The corresponding corner of parameter space is characterized by
light $G^\prime$ masses $M_{G^\prime} \lesssim 300$~GeV.  As the
currently available results for dijet pair searches do not exclude the
range between 200 GeV and 320 GeV, a heavy gluon cannot be ruled out
as a possible NP explanation of the observed $\Delta A_{\rm CP}$.  If
a heavy gluon is indeed responsible for the large value of $\Delta
A_{\rm CP}$, indirect CPV in $D^0 - \bar D^0$ mixing is also expected
to be close to the current experimental bounds.

\subsection{Charged Vector Boson} \label{subsec:Wprime}

We consider a new vector boson with charge $\pm 1$ that couples to
right-handed up and down type quarks
\begin{equation} \label{eqn:Wprime}
\mathcal{L}_{\text{int}} = g_R V^R_{ij} \bar d_R^i  u_R^j 
W^{\prime -} + h.c. ~,
\end{equation}
where $V^R_{ij}$ is a unitary mixing matrix, the analog of the CKM
matrix in the right-handed sector.  One possibility to introduce such
a $W^\prime$ gauge boson is through an additional $SU(2)_R$ gauge
group~\cite{Schmaltz:2010xr}.  Yet as long as the coupling structure
in~\eqnref{Wprime} is realized, the exact implementation of the
$W^\prime$ is of no relevance for the following discussion.

Tree level exchange of the $W^\prime$ gives contributions to the
current-current Wilson coefficient $\tilde C_1^{(1)p}$. The
corresponding operator $\tilde O_1^{(1)p}$ has the same hadronic
matrix element as the SM operator $O_1^{(1)p}$. Therefore, tree level
exchange of the $W^\prime$ cannot generate a direct CP asymmetry
because there is no strong phase difference with respect to the LO SM
contribution.

Loop level contributions, {\it i.e.} gluon penguins with $W^\prime$
loops, have a structure that is analogous to the SM penguin
contribution. If we assume that the mass of the $W^\prime$ is larger
than the mass of the SM $W$ boson, then the couplings $g_R V^R_{cb}
V^{R*}_{ub}$ have to be considerably larger than the SM couplings $g
V_{cb} V_{ub}^*$ in order to generate a sizable $\Delta A_{\rm CP}$.
Such a $W^\prime$ would then lead to unacceptably large NP
contributions to $B \to D$ and $B \to \pi$ transitions, and thus we
will not consider this scenario any further.

\subsection{Two Higgs Doublet Model} \label{subsec:2HDM}
%
\begin{figure*}[tbp]
\centering
\includegraphics[width=0.30\textwidth]{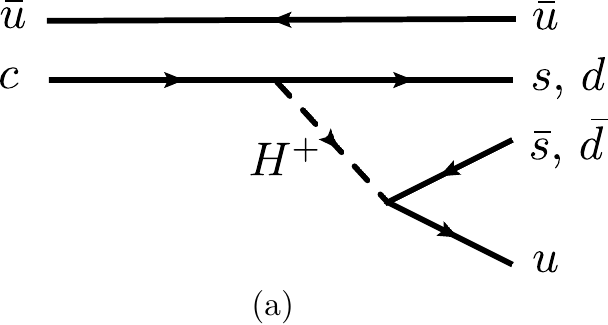} ~~~~~~
\includegraphics[width=0.24\textwidth]{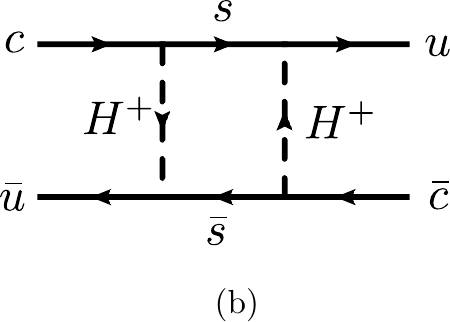} ~~~~~~
\includegraphics[width=0.24\textwidth]{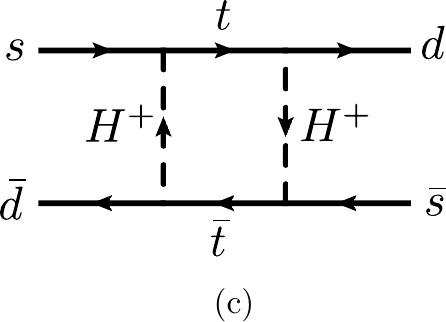} \\[10pt]
\includegraphics[width=0.27\textwidth]{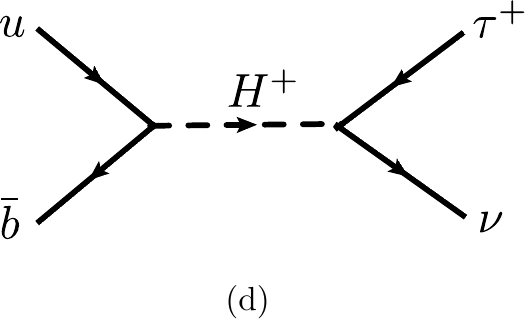} ~~~~~~
\includegraphics[width=0.25\textwidth]{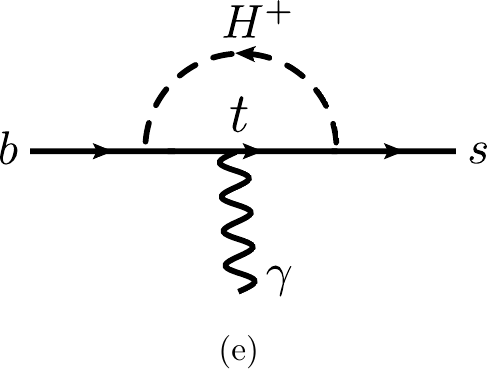}
\caption{Example Feynman diagrams that contribute to~(a) the $D \to
  K^+K^-$ and $D \to \pi^+\pi^-$ decay amplitudes,~(b) $D^0 - \bar
  D^0$ mixing,~(c) $K - \bar K$ mixing,~(d) the $B^+ \to \tau
  \nu$ decay and~(e) the $B_d \to X_s \gamma$ decay in the
  discussed 2HDM with MFV.}
\label{fig:Diagrams_2HDM}
\end{figure*}
%
One of the simplest extensions of the SM scalar sector is the 2 Higgs
doublet model (2HDM) (see~\cite{Branco:2011iw} for a recent
review). The most general couplings of the 2 Higgs bosons to SM
fermions read
\begin{eqnarray} \label{eqn:2HDM}
\mathcal{L}_{\text{int}} &=& Y_u \bar Q U H_u + Y_d \bar Q D H_d + 
Y_\ell \bar L E H_d \\
&+& X_u \bar Q U H_d^\dagger + X_d \bar Q D H_u^\dagger + X_\ell \bar L
E H_u^\dagger + \text{ h.c.} \nonumber ~,
\end{eqnarray}
where we suppress the flavor indices on the quark fields and the
coupling matrices $X$ and $Y$.  Generically, the neutral components of
both Higgs bosons acquire a vev $\langle H^0_{u,d} \rangle = v_{u,d}$
and the fermion mass matrices receive contributions from both $Y$ and
$X$. As $Y$ and $X$ are independent matrices in flavor space, the
couplings in~\eqnref{2HDM} lead to flavor changing neutral Higgs
vertices at tree level and are strongly constrained from meson mixing
observables.

The most effective way to control flavor changing effects in the 2HDM is the
Minimal Flavor Violation (MFV) ansatz~\cite{D'Ambrosio:2002ex,
Buras:2010mh}. Applied to the 2HDM, the MFV assumption states that the
``wrong'' Higgs couplings $X$ can be expanded in powers of the Yukawa
couplings $Y$
\begin{eqnarray}
 X_u &=& \epsilon_u Y_u + \epsilon_u^\prime Y_u Y_u^\dagger Y_u 
+ \epsilon_u^{\prime\prime} Y_d Y_d^\dagger Y_u + \dots ~,\nonumber \\
 X_d &=& \epsilon_d Y_d + \epsilon_d^\prime Y_u Y_u^\dagger Y_d 
+ \epsilon_d^{\prime\prime} Y_d Y_d^\dagger Y_d + \dots ~,
\end{eqnarray}
where the $\epsilon_i$ are free complex parameters.  For simplicity,
we will also assume $X_\ell = \epsilon_\ell Y_\ell$.

We stress that in this particular 2HDM, there exists no preferred
basis for the 2 Higgs doublets. The $\epsilon_i$ parameters as well as
$\tan\beta = v_u/v_d$ are basis dependent and none of them separately
are actually physical parameters (see~\cite{Davidson:2005cw} for a
detailed discussion). In the following analysis, we fix a basis by
setting $\epsilon_d = 0$. In this basis, a large $\tan\beta$ can be
approximately identified with the basis invariant enhancement of the
coupling of the right handed strange quark to the charged Higgs with
respect to its SM Yukawa coupling.

We now investigate the parameter space of this 2HDM with MFV in a
basis with $\epsilon_d = 0$. We work in the regime of large
$\tan\beta$ and assume $\epsilon_u^{\prime\prime}, \epsilon_d^\prime
\lll 1$ in order to ensure that tree level FCNCs are under
control. Furthermore, we allow the parameters $\epsilon_u,
\epsilon_u^\prime$ and $\epsilon_d^{\prime\prime}$ as well as
$\epsilon_\ell$ to be $\mathcal{O}(1)$. As we will see below, in this
region of parameter space, sizable nonstandard contributions to the
$D \to K^+ K^-$ amplitude can arise, while the most important
constraints can be kept under control.

In this scenario, tree level charged Higgs exchange, as shown in
diagram~(a) of~\figref{Diagrams_2HDM}, gives the dominant NP
contribution to the $D \to K^+ K^-$ decay.  For large values of
$\tan\beta$ and assuming $\epsilon_u \sim \mathcal{O}(1)$ we
find\footnote{Note that this expression as well as the ones given
below are not basis invariant. They only hold in bases where the same
conditions on the $\epsilon_i$ and $\tan\beta$ hold, under which they
were derived.}
\begin{equation}
\tilde C_{S1}^{(1)} = \frac{m_c m_s}{v^2} \epsilon_u \frac{\tan\beta}{1 
+ \tilde\epsilon_s \tan\beta} 
\frac{V_{us}V_{cs}^*}{M_{H^\pm}^2} ~,
\end{equation}
with $v^2 = v_u^2 + v_d^2 = 174^2$~GeV$^2$, and $\tilde\epsilon_s
\equiv \epsilon_d^{\prime\prime}y_s^2$. For large $\tan\beta$ and
$\epsilon_d^{\prime\prime} \sim \mathcal{O}(1)$, we find
$\tilde\epsilon_s \sim 10^{-3}$, such that $\tilde\epsilon_s$ only
becomes relevant for extremely large $\tan\beta$. The parameter
$\epsilon_u$ simultaneously lifts the $1/\tan\beta$ suppression of the
$\bar c_R s_L H^+$ vertex and provides a source of CPV.

\bigskip

The 2HDM with MFV have been thoroughly studied in the literature
(see~\cite{Buras:2010mh, Buras:2010zm, Trott:2010iz, Jung:2010ik,
Cline:2011mm}) and various constraints have been identified. In the
following, we discuss the most important constraints:

(i) Direct searches at LEP for a charged Higgs give the bound
$M_{H^\pm} \gtrsim 80$~GeV~\cite{Heister:2002ev}.

(ii) The $B^+ \to \tau^+ \nu$ and $K^+ \to \mu^+ \nu$ decays are known
to be important low energy probes of extended Higgs sectors.
Combining the experimental results from Belle~\cite{Ikado:2006un,
Hara:2010dk} and BaBar~\cite{:2008gx, :2010rt} on Br$(B \to \tau\nu)$
with a conservative SM prediction based on $|V_{ub}| = (3.89 \pm 0.44) \times
10^{-3}$~\cite{Nakamura:2010zzi} and $f_{B^+} = 196.9 \pm
8.9$~MeV~\cite{Bazavov:2011aa}
\begin{equation}
{\rm Br}(B \to \tau\nu)_{\rm SM} = (1.04 \pm 0.25) \times 10^{-4} ~,
\end{equation}
we find
\begin{equation}
R_{B\tau\nu} = \frac{{\rm Br}(B \to \tau\nu)_{\rm exp}}{{\rm Br}(B \to
  \tau\nu)_{\rm SM}} = 1.58 \pm 0.32 ~.
\end{equation}
In our 2HDM, tree level exchange of a charged Higgs (see diagram~(d)
in~\figref{Diagrams_2HDM}) results in
\begin{equation}
R_{B\tau\nu} = \left| 1 - \frac{m_B^2}{M_{H^\pm}^2}
\frac{\tan\beta}{1+\tilde\epsilon_b \tan\beta} \frac{1}{\epsilon_\ell} 
\right|^2~,
\end{equation}
where we defined $\tilde\epsilon_b \equiv \epsilon_d^{\prime\prime} y_b^2$.
For large $\tan\beta$ and $\epsilon_d^{\prime\prime} \sim
\mathcal{O}(1)$, one typically has $\tilde\epsilon_b \sim 10^{-2} -
10^{-1}$.  The factors $\tilde\epsilon_b$ and $\epsilon_\ell$ keep the
$\bar b_R u_L H^\pm$ and $\bar \tau_R \nu_L H^\pm$ couplings small for
large $\tan\beta$ and the experimental constraint from $B^+ \to \tau^+
\nu$ is easily avoided, provided ${\rm sign}(\epsilon_\ell) = -1$.  In
fact, for such choices of parameters, the $\sim 2\sigma$ discrepancy
between the SM prediction and the experimental result for Br$(B \to
\tau\nu)$ is reduced.

For the observable $R_{\ell23}$~\cite{Antonelli:2010yf}, which is
sensitive to charged Higgs contributions to the $K \to \mu \nu$ decay,
we find
\begin{equation}
R_{\ell23} = \left| 1 - \frac{m_K^2}{M_{H^\pm}^2}
\frac{\tan\beta}{1+\tilde\epsilon_s \tan\beta} \frac{1}{\epsilon_\ell}
\right|~.
\end{equation}
From the experimental side, one has~\cite{Antonelli:2010yf}
\begin{equation}
R_{\ell23} = 0.999 \pm 0.007~,
\end{equation}
which, in our framework, leads only to constraints for extremely large
$\tan\beta$.

(iii) Charged Higgs loops lead to contributions to $D^0 - \bar D^0$
and $K - \bar K$ mixing (see diagrams~(b) and~(c)
of~\figref{Diagrams_2HDM}). In the considered scenario, we find the
following dominant NP contributions
\begin{eqnarray}
C_1^{(2) D} &\simeq& \frac{-1}{16\pi^2} \frac{m_s^4}{v^4} 
\frac{(V_{cs}V_{us}^*)^2}{8 M_{H^\pm}^2}
\frac{\tan^4\beta}{|1+\tilde\epsilon_s \tan\beta|^4} ~, \\
C_1^{(2) K} &\simeq& \frac{-1}{16\pi^2} \frac{m_t^4}{v^4} 
\frac{(V_{ts}V_{td}^*)^2}{4 M_{H^\pm}^2} \nonumber \\
&& \Big( |\tilde\epsilon_t|^4 h_1(x_t) + 
|\tilde\epsilon_t|^2 h_2(x_t,x_W) \Big)~,
\end{eqnarray}
with $x_t = m_t^2/M_{H^\pm}^2$, $x_W = M_W^2/M_{H^\pm}^2$,
$\tilde\epsilon_t \equiv \epsilon_u + \epsilon_u^\prime y_t^2$, and the
analytical expressions for the loop functions $h_1$ and $h_2$ are
given in~\appref{loop_functions}. The expression for $C_1^{(2) D}$ is
suppressed by four powers of the strange quark mass and is only
relevant for very large $\tan\beta$.\footnote{The corresponding
contribution from a bottom quark loop is strongly Cabibbo suppressed
and turns out to be much smaller.} The Wilson coefficient $C_1^{(2)
K}$ arises from box diagrams including both one $W^\pm$ and one
$H^\pm$ insertion as well as diagrams with two $H^\pm$ insertions. Its
contribution to kaon mixing can be relevant for $\tilde\epsilon_t
\simeq \mathcal{O}(1)$. Yet even for $|\tilde\epsilon_t| \simeq 0.5$,
the constraint on the charged Higgs mass is as low as the bound from
direct searches $M_{H^\pm} \gtrsim 80$~GeV.

(iv) As shown in diagram~(e) of~\figref{Diagrams_2HDM}, the $B_d \to
X_s \gamma$ decay also receives 1-loop charged Higgs contributions. As
is well known, the good agreement of the experimental data and the SM
prediction of its branching ratio leads to the constraint $M_{H^\pm}
\gtrsim 300$~GeV in a 2HDM of type~II~\cite{Misiak:2006zs}. This
bound, however, does not apply in the model considered here. For the
ratio of the $b \to s \gamma$ amplitudes in our 2HDM with MFV model to
the 2HDM of type~II, we find, to a good approximation,
\begin{equation}
\frac{A(b\to s\gamma)_{\rm MFV}}{A(b\to s\gamma)_{\rm II}} =
\frac{\tilde\epsilon_t \tan\beta}{1 + \tilde\epsilon_b \tan\beta} +
|\tilde\epsilon_t|^2 y(x_t)~,
\end{equation}
where $y(x_t)$ is a function of $x_t = m_t^2 / M_{H^\pm}$ and is
$\mathcal{O}(1)$. We point out that because of the $\tilde\epsilon_t$
and $\tilde\epsilon_b$ factors, the $b \to s \gamma$ amplitude can be
complex, which relaxes the $B_d \to X_s \gamma$ constraint
considerably~\cite{Altmannshofer:2011gn}. Still, charged Higgs masses
as low as the constraint from direct searches require
$\tilde\epsilon_t = \epsilon_u + \epsilon_u^\prime y_t^2 \ll 1$, which
implies a considerable amount of fine tuning.

(v) Complex $\epsilon_i$ parameters also lead to contributions to
electric dipole moments (EDMs) (see~\cite{Buras:2010zm,
Trott:2010iz}). While a detailed study of EDMs is beyond the scope of
this work, we mention that in the studied framework, the contributions
to the EDMs depend on the parameters $\tilde\epsilon_t$ and
$\tilde\epsilon_b$. Therefore, by allowing for a certain amount of
fine tuning, EDM constraints do not exclude sizable CP violating
effects in the $D \to K^+ K^-$ decay that depend mainly on
$\epsilon_u$.

(vi) Charged Higgses are also constrained by bounds on the branching
ratio of the top quark decay $t \to H^+ b$.  ATLAS and CMS obtained
bounds at the level of Br$(t \to H^+ b) \lesssim 5\%$ under the
assumption Br$(H^- \to \tau \nu) = 1$~\cite{HpmLHC}. CDF and D\O\ also
give bounds considering the $H \to c s$ final state. These bounds are
at the level of Br$(t \to H^+ b) \lesssim 10\% -
20\%$~\cite{Aaltonen:2009ke,:2009zh}.  In our setup, the ratio of the
$t \to H^+ b$ and $t \to W b$ branching ratios is given by
\begin{eqnarray}
&& \frac{\Gamma_{tHb}}{\Gamma_{tWb}} = 
\frac{1-(m_H/m_t)^2}{1-(m_W/m_t)^2} 
\frac{1}{m_W^2m_t^2 + m_t^4 - 2m_W^4} \nonumber \\
&& ~~ \times \left[ \left( m_t^2 |\tilde\epsilon_t|^2 + 
\frac{m_b^2 \tan^2\beta}{|1+\tilde\epsilon_b \tan\beta|^2} \right) 
(m_t^2 - m_H^2) \right. \nonumber \\
&& ~~\left. + 4 {\rm Re} 
\left(\frac{\tilde\epsilon_t \tan\beta}{1 + \tilde\epsilon_b \tan\beta}\right) 
m_t^2 m_b^2\right] ~.
\end{eqnarray}
For the parameter choices detailed below, we find the charged Higgs
branching ratio into $\tau \nu$ does not exceed $10\%$ due to the
strongly enhanced couplings to strange quarks. Also, since Br$(t \to
H^\pm b) \lesssim 10\%$, top decays do not lead to constraints for our
choice of parameters.

(vii) Additional constraints on the 2HDM can come from direct searches
for neutral Higgs bosons. The relation between the charged and the
neutral Higgs bosons, however, depends on the details of the Higgs
potential. For simplicity we assume a Higgs potential such that there
is one neutral Higgs boson $h$ that has SM-like couplings to gauge
bosons and fermions. Consequently, the other two bosons $H$ and $A$ do
not couple to gauge bosons and have couplings to bottom quarks that
are enhanced by $\tan\beta/(1+\tilde\epsilon_b \tan\beta)$. While the
masses of the Higgs bosons are in principle free parameters, we will
assume $M_H \simeq M_A \simeq M_{H^\pm}$ to avoid constraints from
electroweak precision observables.  Constraints can arise from neutral
Higgs boson searches in the $H \to \tau \tau$ final state at ATLAS and
CMS~\cite{H0LHC}, as well as from searches at D\O\ and CDF in the $H
\to b b$ final state~\cite{Abazov:2010ci,CDF_3b}.

\begin{figure}[tbp]
\centering
\includegraphics[width=0.45\textwidth]{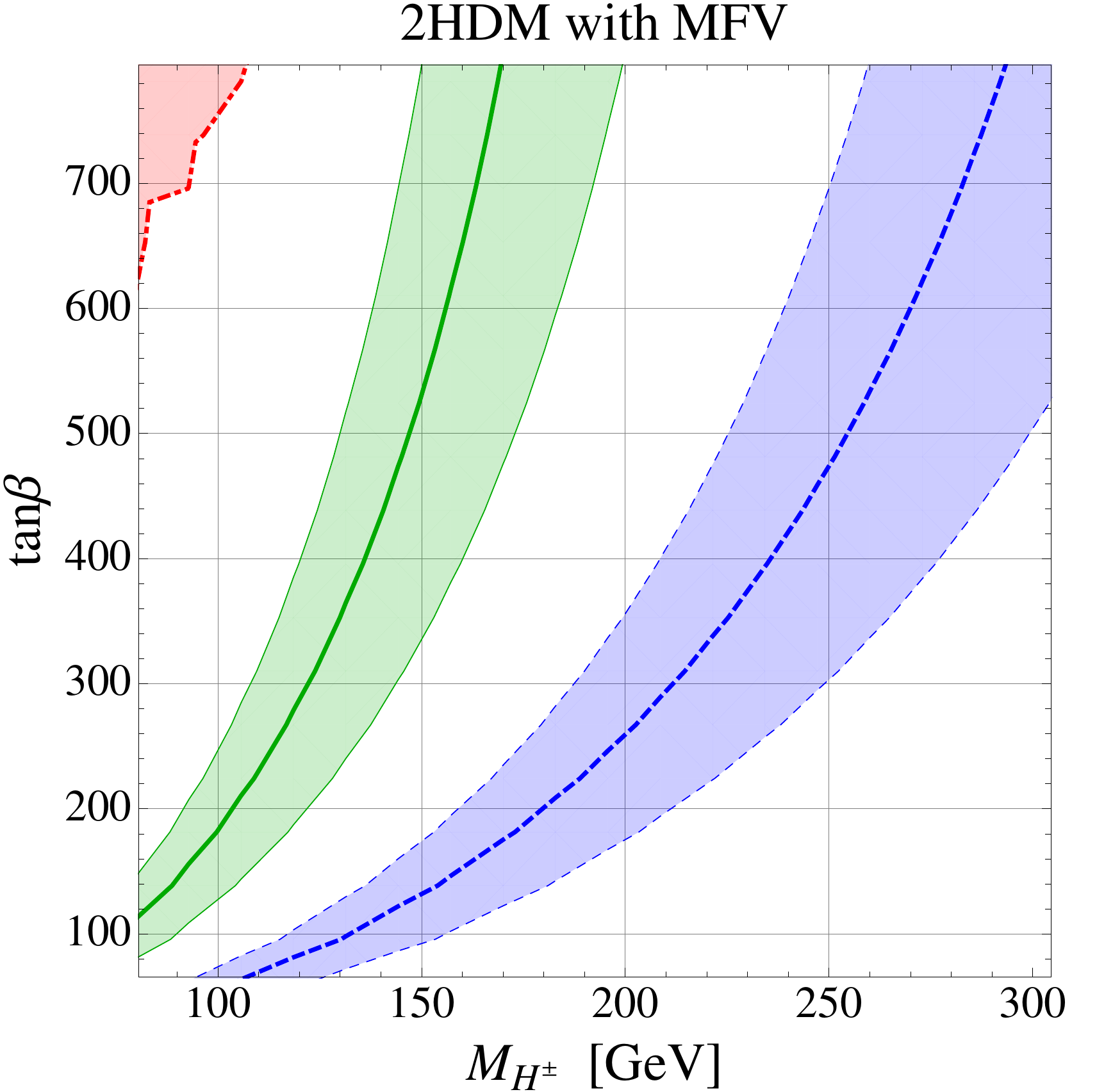}
\caption{Regions in the $M_{H^\pm}$ - $\tan\beta$ plane compatible
  with the data on $\Delta A_\text{CP, WA}$ at the $1\sigma$ level in the
  discussed 2HDM. In a basis with $\epsilon_d = 0$, we set $\epsilon_u
  = i$, $\tilde{\epsilon}_t = 0.05i$, $\epsilon_\ell = -1$ and chose
  $\epsilon_d^{\prime \prime}$ such that $\tilde\epsilon_s =
  10^{-3}$. All other $\epsilon_i$ are set to zero. The green (solid)
  band corresponds to the expressions for the decay amplitude in
  na\"{i}ve factorization, blue (dashed) band assumes an enhancement
  of the hadronic matrix elements by a factor of 3. The red
  (dash-dotted) region is excluded by $D^0 - \bar D^0$ mixing
  constraints.}
\label{fig:2HDM}
\end{figure}

In the following, we discuss a benchmark scenario that avoids all
considered constraints but allows for nonstandard values for $\Delta
A_{\rm CP}$. We set $\epsilon_u = i$, $\tilde{\epsilon}_t = 0.05i$,
$\epsilon_\ell = -1$, and chose $\epsilon_d^{\prime \prime}$ such that
$\tilde\epsilon_s = 10^{-3}$. All other $\epsilon_i$ are set to zero.
\figref{2HDM} shows the regions in the $M_{H^\pm}$ -- $\tan\beta$
plane compatible with the data for $\Delta A_\text{CP, WA}$ at the
$1\sigma$ level. The green (solid) band is obtained using the
expressions for the decay amplitude in na\"{i}ve factorization. The
blue (dashed) band assumes an enhancement of the hadronic matrix
elements by a factor of 3. The red (dash-dotted) region in the upper
left corner is excluded by $D^0 - \bar D^0$ mixing constraints.

We observe that sizable NP contributions to the direct CP asymmetry
are only possible for small charged Higgs masses and large values of
$\tan\beta \sim \mathcal{O}(100)$ and larger. It is important to note
that such large values for $\tan\beta$ are not in conflict with a
requirement of perturbative Yukawa couplings. Indeed, due to the
$\tilde\epsilon_b$ (we find $0.03 < \tilde\epsilon_b < 0.09$ in the
considered region of $\tan\beta$) and $\epsilon_\ell$ factors, the
bottom and tau Yukawa remain well below $1$.~\footnote{The
phenomenology of generating appropriate bottom and tau masses in the
very large $\tan\beta$ regime in the context of supersymmetric models
was studied in~\cite{Dobrescu:2010mk, Altmannshofer:2010zt}. We remark
however that the considered scenario cannot be realized in the MSSM
where the $\epsilon_i$ factors are only loop induced and not much
larger than $10^{-2}$.}  Extremely large values of $\tan\beta$ are not
stable under radiative corrections. Generically, 1-loop corrections to
the Higgs potential lead to $1/\tan\beta \sim 1/(4\pi)^2$. Values of
$\tan\beta \simeq {\rm few} \times 100$, however, seem a reasonable
possibility.
 
The shown region of parameter space is not significantly constrained
by any of the considered bounds. In particular, the bound from Br$(B_d
\to X_s \gamma)$ is avoided because of the small $\tilde\epsilon_t$.
As stated before, the branching ratio of the charged Higgs into $\tau
\nu$ does not exceed $10\%$ due to the strongly enhanced couplings to
strange quarks, and in addition, constraints from top decays are
satisfied since Br$(t \to H^\pm b) \lesssim 10\%$. Similarly, we find
that the branching ratios of the neutral Higgs bosons to $\tau \tau$
are tiny, $\mathcal{O}(0.1\%)$, and the stringent bounds from ATLAS
and CMS~\cite{H0LHC} are easily avoided. Finally, due to the
$\tilde\epsilon_b$ factor that controls the size of the $\bar b b H$
and $\bar b b A$ couplings, the $H \to bb$ searches at
Tevatron~\cite{Abazov:2010ci,CDF_3b} do not lead to constraints.  We
expect that the collider phenomenology of this large parameter space
will focus on a combination of complementary probes, including tests
for non-SM Yukawa couplings of the light Higgs boson and heavy flavor
searches for pairs of dijet resonances.  A complete phenomenological
study of this 2HDM with MFV is left for future work.

We stress that the 2HDM with MFV can only significantly affect the
direct CP asymmetry in $D \to K^+ K^-$.  New Physics effects in
$A_{\rm CP}(\pi^+\pi^-)$ are strongly suppressed by $m_d/m_s$ and
negligibly small. The considered model also does not lead to large
nonstandard effects in indirect CPV in $D^0 - \bar D^0$ mixing. These
statements are not necessarily true if the couplings of the Higgses
radically departure from the MFV ansatz (see~\cite{Hochberg:2011ru}
for a related study). Even though such models will generically have
flavor changing neutral Higgs interactions, they can in principle be
made compatible with low energy flavor constraints. A detailed study
of such setups, however, is beyond the scope of this work.

\subsection{Scalar Octet} \label{subsec:octet}

\begin{figure}[tbp]
\centering
\includegraphics[width=0.23\textwidth]{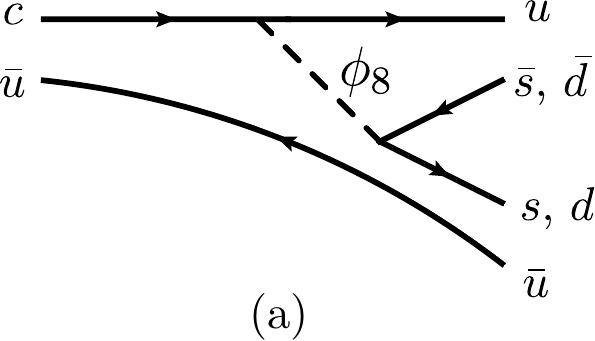} ~~~
\includegraphics[width=0.20\textwidth]{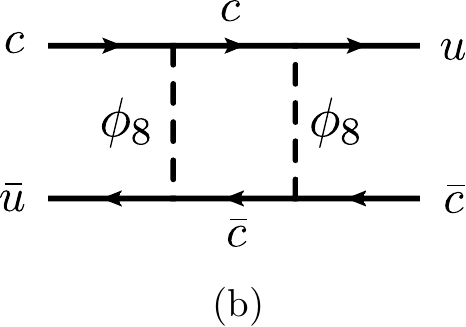}
\caption{Example Feynman diagrams that contribute to~(a) the $D \to
  K^+K^-$ and $D \to \pi^+\pi^-$ decay amplitudes and~(b) $D^0 - \bar
  D^0$ mixing in the scalar octet model.}
\label{fig:Diagrams_octet}
\end{figure}

Couplings analogous to the ones in~\eqnref{2HDM} are also possible if
the second scalar $SU(2)_L$ doublet is a color octet instead of a color
singlet~\cite{Manohar:2006ga}.  We write
\begin{eqnarray} \label{eqn:octet}
\mathcal{L}_{\text{int}} &=& Y_u \bar Q U H + 
Y_d \bar Q D H^\dagger \\
&+& X_u \bar Q T^a U \phi_8^a + X_d \bar Q T^a D \phi_8^{a \dagger} 
+ \text{ h.c.} \nonumber
\end{eqnarray}
where $H$ is the SM Higgs and $\phi_8^a$, $a = 1, \ldots 8$, is the
color octet. Since a color octet must not aquire a vev, the quark
masses are entirely provided by the SM Higgs, and the Yukawa couplings
$Y$ are given by their SM values.  As in the case of the 2HDM, we will
first assume that the couplings $X$ have the MFV structure
\begin{eqnarray}
 X_u &=& \zeta_u Y_u + \zeta_u^\prime Y_u Y_u^\dagger Y_u ~,\nonumber \\
 X_d &=& \zeta_d Y_d + \zeta_d^\prime Y_d Y_u^\dagger Y_u ~.
\end{eqnarray}
We can neglect higher powers of the down-type Yukawa couplings in the
expansion because in the considered scenario they are fixed to their
small SM values. This automatically ensures that there are no tree
level flavor changing interactions of the neutral component of the
scalar octet with up quarks.  Furthermore, the interactions of the
charged component of the scalar octet cannot lead to tree level
contributions to the $D \to K^+ K^-$ amplitude in na\"{i}ve
factorization, as the two quarks that couple to the color octet cannot
hadronize into a kaon or pion, which are color singlets.
Contributions to the $D \to K^+ K^-$ decay can, however, come from
annihilation topologies involving the charged octet state. Even though
such contributions are formally suppressed by $1/m_c$, they can be
sizable in $D$ meson decays.  Nonetheless, we now consider departures
from the MFV ansatz for the couplings and come back to the MFV
framework at the end of the section.  In particular, we consider the
following (small) non-MFV entry in the $X_u$ coupling matrix in the
basis where the quark Yukawa couplings are diagonal
\begin{equation}
 \left[ X_u \right]_{12} = \zeta_u y_c X_{cu} ~,
\end{equation}
with $X_{cu} \sim V_{cs} V_{us}^* \sim \lambda \sim 0.2$.  In contrast
to the MFV interactions, the above term leads to a flavor changing
neutral current $u_L c_R \phi_8^0$ coupling and, due to $SU(2)_L$
invariance, to a $\mathcal{O}(1)$ correction of the $d_L c_R
\phi_8^\pm$ coupling
\begin{eqnarray}
\mathcal{L}_{\text{int}} &=& \zeta_u y_c X_{cu} ~ \bar u_L T^a c_R 
~ \phi_8^{0a} \nonumber \\
&+& \zeta_u y_c (V_{cd}^* + X_{cu} V_{ud}^*) ~\bar d_L T^a c_R 
~ \phi_8^{- a} + \text{ h.c.} ~.
\end{eqnarray}

The tree level exchange of the neutral component of the scalar octet,
as shown in diagram (a) of~\figref{Diagrams_octet}, then leads to the
following contributions to the Wilson coefficients,
\begin{eqnarray}
\tilde C_{S1}^{(1)} &=& \frac{m_c m_s}{v^2} \zeta_u \zeta_d
\frac{X_{cu}}{4 M_{\phi_8}^2} ~, \\
\tilde C_{S2}^{(1)} &=& - \frac{4}{N_c} \tilde C_{T1}^{(1)} = 4 \tilde C_{T2}^{(1)} = 
-\frac{1}{N_c} \tilde C_{S1}^{(1)} ~.\nonumber 
\end{eqnarray}
Even though the tensor operators do not contribute to the $D \to K^+
K^-$ decay, we include the corresponding Wilson coefficients here
because they mix with the scalar operators under renormalization.

Tree level contributions to $D^0 - \bar D^0$ mixing from the exchange
of the neutral {\it complex} scalar $\phi_8^0$ can only arise if both
the flavor changing couplings $\bar u_L c_R \phi_8^0$ and $\bar c_L
u_R \phi_8^0$ are present simultaneously. Thus, in our setup,
contributions to $D^0 - \bar D^0$ mixing first arise at the loop
level. The dominant contribution comes from a charm quark loop, as
shown in diagram (b) of~\figref{Diagrams_octet},
\begin{equation}
C_1^{(2)D} = \frac{1}{16\pi^2} \frac{N_c^3 - 2 N_c +1}{4N_c^2}
\frac{m_c^4}{v^4} \frac{X_{cu}^2}{8M_{\phi_8}^2} |\zeta_u|^4 ~.
\end{equation}

Similar to the situation discussed in the context of the 2HDM above,
other flavor changing processes like kaon mixing or $B_d \to X_s
\gamma$ do not directly constrain $\zeta_u$ or $\zeta_d$ as they
involve couplings to right-handed top and bottom quarks that are
proportional to $\zeta_t = \zeta_u + \zeta_u^\prime y_t^2$ and
$\zeta_b = \zeta_d + \zeta_d^\prime y_t^2$, respectively. Keeping
these couplings small to avoid constraints from perturbativity and, in
particular from $B_d \to X_s \gamma$, requires a considerable amount
of fine tuning.  On the other hand, since color octets cannot couple
to leptons, the constraints from $B \to \tau \nu$ and $K \to \mu \nu$
are automatically avoided.

Constraints from electroweak precision observables are under control,
provided all components of the scalar octet have approximately the
same mass and are heavier than $100$~GeV~\cite{Burgess:2009wm}.

There has been extensive work done on collider constraints on scalar
octets~\cite{Gresham:2007ri, Bai:2010dj}. Because of the small
couplings to light quarks in the considered framework, single
production of scalar octets is strongly suppressed, and dijet
resonance searches at the hadron colliders give no significant
constraints in the considered region of parameter space.  Instead, the
main collider constraints on the scalar octet doublet come from the
dijet pair searches at the LHC. We simulate the production of scalar
octet pairs (including production of neutral pairs and charged pairs)
using \verb|MadGraph 5| and compare the obtained cross sections with
the bounds set by ATLAS~\cite{Aad:2011yh} and
CMS~\cite{CMS-EXO-11-016}.  We find that the region from 100~GeV to
200~GeV that is covered by the ATLAS search is fully excluded. The CMS
search excludes scalar octet masses from 320~GeV until approximately
550~GeV for 13\% acceptance (see~\subsecref{Gprime} for comments
regarding the 200--320 GeV region).  As our scalar octets decay
predominantly to $b$ quarks, searches for final states with multiple
$b$ jets at Tevatron can potentially lead to constraints. Using the
CDF result~\cite{CDF_3b_old}, the authors of~\cite{Gerbush:2007fe}
find that scalar octet masses of $M_{\phi_8} \lesssim 200$~GeV are
excluded. Updated results from CDF~\cite{CDF_3b} and D\O\
~\cite{Abazov:2010ci} do not give significantly improved bounds on the
corresponding cross sections.

\begin{figure}[tbp]
\centering
\includegraphics[width=0.45\textwidth]{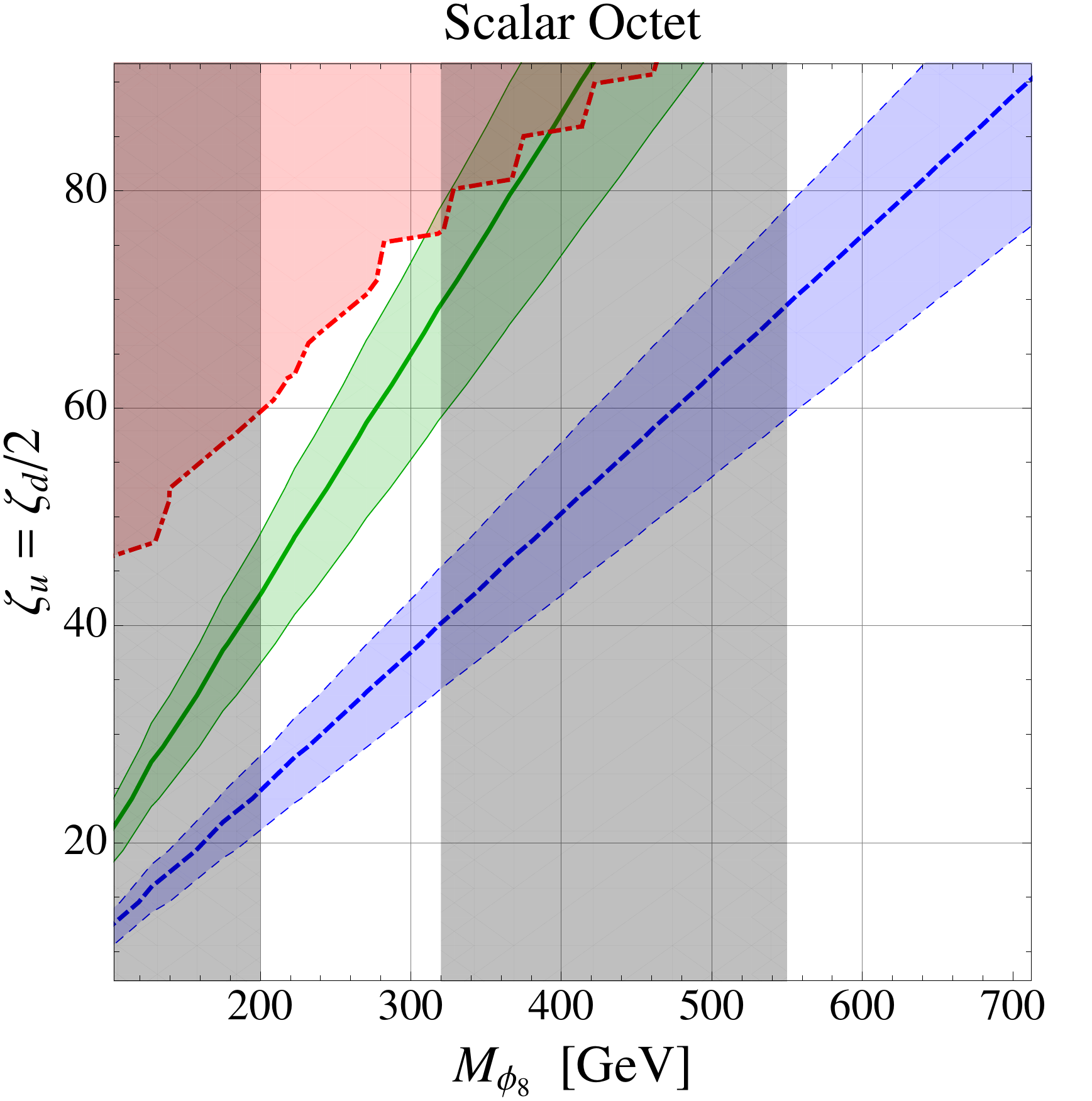}
\caption{Regions in the $M_{\phi_8}$ - $\zeta_u$ plane compatible with
  the data on $\Delta A_\text{CP, WA}$ at the $1\sigma$ level in the scalar
  octet model, assuming $\zeta_u = \zeta_d/2$ real and setting $X_{uc}
  = |V_{cs} V_{us}^*| e^{i \pi/2}$. The green (solid) band corresponds
  to the expressions for the decay amplitude obtained in na\"{i}ve
  factorization, the blue (dashed) band assumes an enhancement of the
  hadronic matrix elements by a factor of 3. The red (dash-dotted)
  region is excluded by $D^0 - \bar D^0$ mixing constraints. The
  vertical shaded bands are excluded by dijet pair searches at LHC.}
\label{fig:octet}
\end{figure} 

In~\figref{octet}, we show the regions in the $M_{\phi_8}$ - $\zeta_u$
plane that are compatible with the data for $\Delta A_\text{CP, WA}$ at the
$1\sigma$ level, assuming $\zeta_u = \zeta_d/2$ real and setting
$X_{cu} = |V_{cs} V_{us}^*| e^{i\pi/2}$. The green (solid) band
corresponds to the expressions for the decay amplitude obtained in
na\"{i}ve factorization, and the blue (dashed) band assumes an
enhancement of the hadronic matrix elements by a factor of 3. The red
(dash-dotted) region is excluded by $D^0 - \bar D^0$ mixing
constraints and the vertical bands are excluded by the dijet pair
searches at ATLAS (left band) and CMS (right band).  As expected,
constraints from $D^0 - \bar D^0$ mixing allow for a sizable $\Delta
A_{\rm CP}$ only for small scalar masses.  Given the constraints from
the dijet pair searches at LHC, only a small window with octet masses
between 200~GeV and 320~GeV and large values for $\zeta_{u,d} \gtrsim
20$ remains where significant values for $\Delta A_{\rm CP}$ can be
generated by the scalar octet with couplings close to the MFV ansatz
without assuming any enhancement of the hadronic matrix
elements. Scalar octets heavier than the CMS bound of $M_{\phi_8}
\simeq$ 550~GeV can lead to nonstandard values for $\Delta A_{\rm
CP}$ only for extremely large $\zeta_{u,d} \gtrsim 60$ and assuming
the hadronic matrix elements to be 3 times the na\"{i}ve factorization
estimate. The same conclusion holds for the pure MFV setup discussed
at the beginning of the section, provided that the annihilation
contribution with the charged scalar exchange is of similar size as
the na\"{i}ve factorization contribution from the flavor changing
neutral scalar.

Both in the MFV and the quasi-MFV setup, the scalar octet can only
significantly affect the direct CP asymmetry in $D \to K^+ K^-$. New
Physics effects in $A_{\rm CP}(\pi^+\pi^-)$ are strongly suppressed by
$m_d/m_s$ and negligibly small. As in the case of the 2HDM, this is
not necessarily the case if the octet couplings are allowed to deviate
more radically from MFV.  We expect collider searches in the multi-$b$
jet final state, as discussed in~\cite{Gresham:2007ri, Bai:2010dj}, to
provide the best sensitivity to this model.

\subsection{Scalar Diquarks} \label{subsec:diquark}
%
\begin{figure*}[tbp]
\centering
\includegraphics[width=0.30\textwidth]{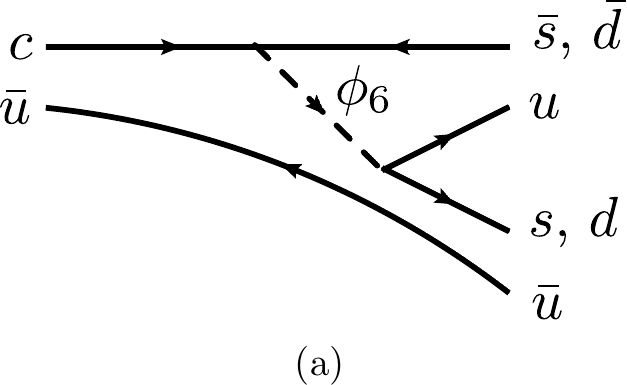} ~~~~~
\includegraphics[width=0.25\textwidth]{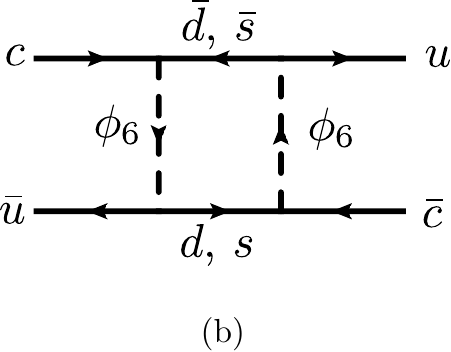} ~~~~~
\includegraphics[width=0.25\textwidth]{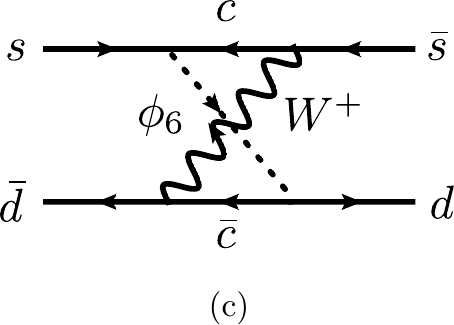}
\caption{Example Feynman diagrams that contribute to~(a) the $D \to
  K^+K^-$ and $D \to \pi^+\pi^-$ decay amplitudes,~(b) $D^0 - \bar
  D^0$ mixing and~(c) $K - \bar K$ mixing in the discussed diquark
  model.}
\label{fig:Diagrams_sextet}
\end{figure*}
%
Numerous variations of scalar diquarks can have renormalizable
couplings to the SM quarks and lead to an interesting flavor
phenomenology~\cite{Arnold:2009ay,Giudice:2011ak}.  Here, we consider
a diquark that has quantum numbers such that it leads to contributions
to meson mixing only at the loop level, but contributes to the $D \to
K^+K^-$ and $D \to \pi^+\pi^-$ decay amplitudes at tree level in
na\"{i}ve factorization.  One such possibility is a scalar that
transforms as $\bar 6$ under $SU(3)$, as a singlet under $SU(2)_L$,
and has hypercharge $-1/3$.~\footnote{All other possibilities are
scalars that can also have lepton number violating couplings. One
example is a $SU(3)$ triplet, $SU(2)_L$ singlet with hypercharge
$-1/3$ which corresponds to a right-handed down squark with $R$-parity
violating couplings.}  Such a diquark can couple to right-handed up
and down type quarks,
\begin{equation}
\mathcal{L}_{\text{int}} = X_{ij} \bar U_i^{c \alpha} D_j^\beta
\phi_6^{\alpha\beta} + \text{ h.c.} ~,
\end{equation}
where $i,j$ are flavor indices and the diquark $\phi_6$ is symmetric
in the color indices $\phi_6^{\alpha\beta} = \phi_6^{\beta \alpha}$.
The couplings $X_{ij}$ are a source of flavor violation and in general
free complex parameters. We remark that the considered diquark can
also couple to left-handed quarks.  Considering couplings to left- and
right-handed quarks, however, simultaneously induces left-right mixing
$\Delta F = 2$ operators, resulting in very severe constraints from
$D$ and $K$ meson mixing. Therefore, we concentrate on couplings to
the right-handed quarks only.

The tree level contributions of the diquark to the $D \to K^+ K^-$ and
$D \to \pi^+ \pi^-$ decay amplitudes are shown in diagram~(a)
of~\figref{Diagrams_sextet} and read
\begin{equation}
\tilde C_1^{(1) p} = \tilde C_2^{(1) p} = \frac{X_{cp} X_{up}^*}{16
  M_{\phi_6}^2} ~.
\end{equation}
Since QCD is parity conserving, the matrix element of the operators
$\tilde O_1^{(1) p}$ are the same as the SM operators $O_1^{(1) p}$,
and no strong phase difference occurs between them.  Correspondingly,
$\tilde C_1^{(1) p}$ does not contribute to the direct CP asymmetries
(see~\eqnref{rf}). Due to the different color structure of the
operators $\tilde O_2^{(1) p}$, however, a different strong phase can
be expected in their matrix elements and a non-zero contribution to
the direct CP asymmetries can be generated by the weak phase in
$\tilde C_2^{(1) p} \propto X_{cp} X_{up}^*$.

Contributions to $D^0 - \bar D^0$ mixing are first induced at the loop
level (see diagram~(b) in~\figref{Diagrams_sextet})
\begin{equation} \label{eqn:DDbar_diquark}
\tilde{C}_1^{(2) D} \simeq - \frac{1}{16 \pi^2} \frac{N_c + 3}{32
  M_{\phi_6}^2} \left( \sum_{p=d,s} X_{up}^*X_{cp} \right)^2 ~,
\end{equation}
where, for simplicity, we restrict ourselves to couplings of the
diquark to the first two generations of quarks and expand the result
in small quark masses.

We note that even with small couplings to quarks $\lesssim 0.1$,
sizable effects in flavor observables can be generated. Consequently,
dijet searches at hadron colliders do not lead to relevant
constraints. Important collider constraints on diquarks, however, come
from the dijet pair searches at LHC. Using \verb|MadGraph 5| to simulate
diquark pair production, we find that the ATLAS search excludes
the region from 100~GeV to 200~GeV. The CMS search excludes diquarks
with masses from 320~GeV until 1000~GeV assuming an acceptance of
13\%.

\begin{figure}[tbp]
\centering
\includegraphics[width=0.45\textwidth]{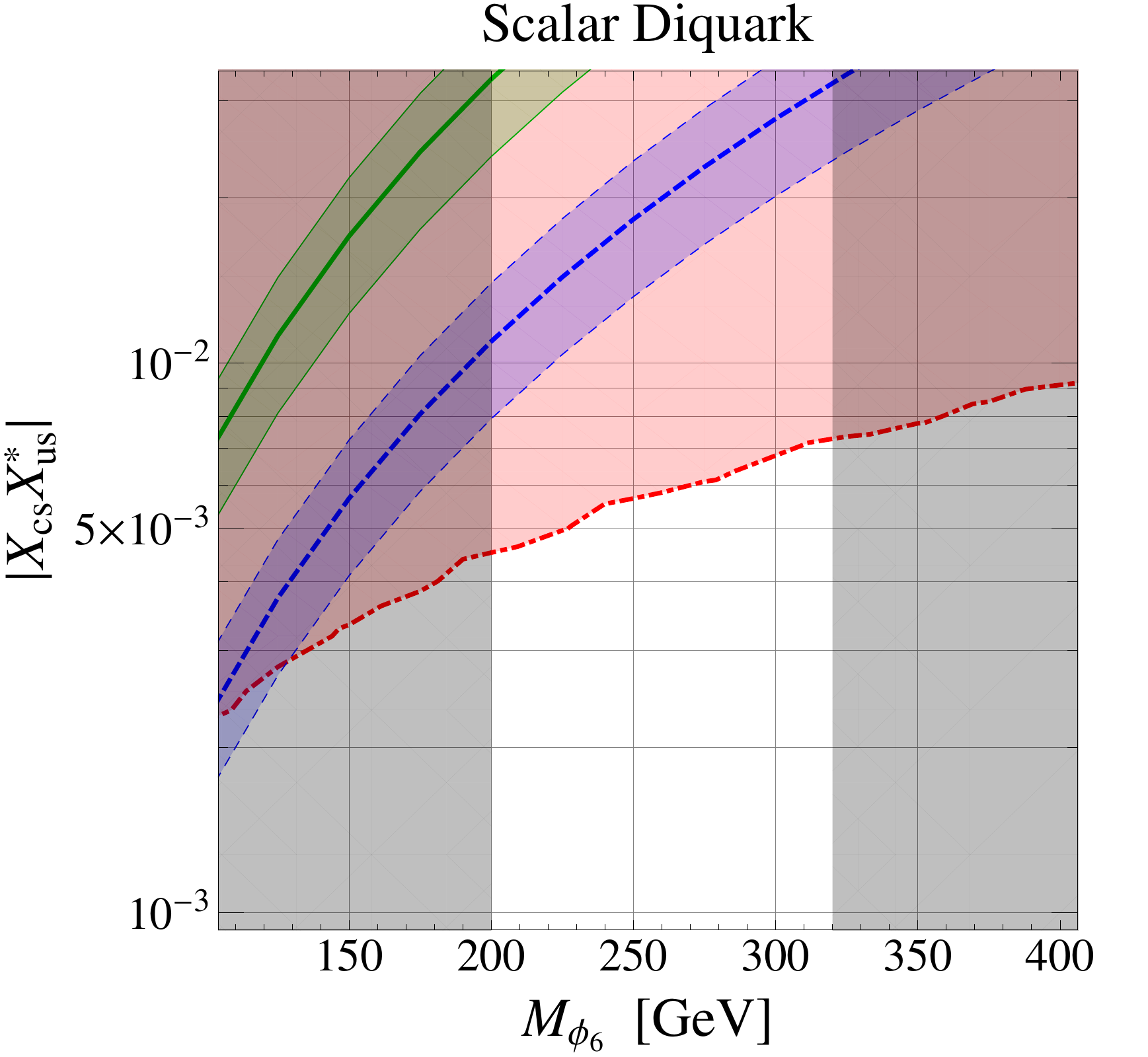}
\caption{Regions in the $M_{\phi_6}$ - $|X_{cs}X_{us}^*|$ plane
  compatible with the data on $\Delta A_\text{CP, WA}$ at the $1\sigma$
  level in the diquark model, setting Arg$(X_{cs}X_{us}^*) =
  \pi/2$. The green (solid) band corresponds to the expressions for
  the decay amplitude obtained in na\"{i}ve factorization, the blue
  (dashed) band assumes an enhancement of the hadronic matrix elements
  by a factor of 3.  The red (dash-dotted) region is excluded by $D^0
  - \bar D^0$ constraints.  The vertical shaded bands are excluded
  by dijet pair searches at LHC.}
\label{fig:diquark}
\end{figure}

In~\figref{diquark}, we show regions in the $M_{\phi_6}$ --
$|X_{cs}X_{us}^*|$ plane compatible with the data for 
$\Delta A_\text{CP, WA}$ at the $1\sigma$ level, considering only $X_{cs}$ and $X_{us}$ to
be non-zero and setting Arg$(X_{cs}X_{us}^*) = \pi/2$. The green
(solid) band corresponds to the expression for the decay amplitude in
na\"{i}ve factorization. The blue (dashed) band assumes an enhancement
of the hadronic matrix elements by a factor of 3. The red
(dash-dotted) region is excluded by the $D^0 - \bar D^0$ mixing
constraints. The vertical bands are excluded by the dijet pair
searches at ATLAS (left band) and CMS (right band). Since only strange
quark couplings are considered, $\Delta A_{\rm CP}$ is entirely
generated by $A_{\rm CP}(K^+K^-)$ in this setup.  Despite the fact
that the $D \to K^+ K^-$ decay amplitude arises already at the tree
level while $D^0 - \bar D^0$ mixing is only induced at the loop level,
constraints from $D^0 - \bar D^0$ mixing combined with the dijet pair
searches exclude a sizable $\Delta A_{\rm CP}$ unless the hadronic
matrix elements are enhanced by a factor more than 3. This is due to
the fact that the contributions to $\Delta A_{\rm CP}$ coming from the
operator $\tilde O_2^{(1) s}$ are color suppressed and neither
enhanced by RG effects nor by chiral factors.  Considering only the
couplings $X_{cd}$ and $X_{ud}$ leads to an analogous situation, but
in this case $\Delta A_{\rm CP}$ entirely stems from $A_{\rm
CP}(\pi^+\pi^-)$.

If all 4 couplings $X_{cs}$, $X_{us}$, $X_{cd}$ and $X_{ud}$ are
present, the $D^0 - \bar D^0$ mixing constraint can be strongly
reduced by assuming a GIM-like mechanism, {\it i.e.} assuming the
coupling matrix $X$ to be (approximately) unitary. This happens, for
example, if $X$ is (approximately) proportional to unity in the weak
eigenstate basis for the quarks. In that case, however, $A_{\rm
CP}(\pi^+\pi^-) \simeq A_{\rm CP}(K^+K^-)$ and the difference between
the two, $\Delta A_{\rm CP}$, is very small.

In addition, also loop contributions to kaon mixing are induced if all
four couplings are present. Apart from a contribution analogous
to~\eqnref{DDbar_diquark}, mixed $\phi_6$ -- $W$ loops (shown in
diagram~(c) of~\figref{Diagrams_sextet}) can become very
important. For the latter, we find
\begin{eqnarray}
C_4^{(2) K} &=& C_5^{(2) K} = \frac{g^2}{16 \pi^2} \left(
V_{cs} V_{cd}^*\right) \left( X_{cs} X_{cd}^*\right) \nonumber \\
&\times& \frac{1}{2 M_{\phi_6}^2} \frac{m_c^2}{M_W^2} \log\left(
\frac{m_c^2}{M_{\phi_6}^2} \right) ~,
\end{eqnarray}
where we expand the result in the charm quark mass and keep only the
leading term that is enhanced by a large logarithm.  Despite the
suppression by $m_c^2/M_W^2$, these contributions are very relevant
because the matrix elements of $O_4^{(2) K}$ and $O_5^{(2) K}$ are
strongly chirally enhanced and enhanced by renormalization group
effects.\footnote{The analogous contribution to $D^0 - \bar D^0$
mixing is suppressed by the strange quark mass and has a much smaller
chiral enhancement from the matrix elements: it is therefore
negligible.}  Furthermore, the simultaneous presence of $X_{us}$ and
$X_{ud}$ also leads to tree level contributions to
$\epsilon^\prime/\epsilon$.

Even after tuning phases in order to minimize constraints from CPV in
kaon mixing and $\epsilon^\prime/\epsilon$, we do not find any region
in parameter space of the considered diquark model in which a sizable
$\Delta A_{\rm CP}$ is compatible with all constraints.

\section{New Physics Contributions to Gluon Penguins}
\label{sec:penguin}
%
We now consider possibilities for new physics in the gluon penguin
diagrams at the 1-loop level. We consider models with new fermion and
scalar fields, where the fermion is Dirac or Majorana and with or
without GIM suppression in~\subsecref{loop_no_GIM}
and~\subsecref{loop_GIM}, respectively.  A scenario with chirally
enhanced magnetic penguins is considered in~\subsecref{chiral}.  As
discussed in the Introduction, the chiral enhancement of the decay
amplitudes allows such a scenario to be the least constrained by $D^0
- \bar D^0$ mixing.

\begin{figure*}[tbp]
\centering
\raisebox{13pt}{\includegraphics[width=0.30\textwidth]{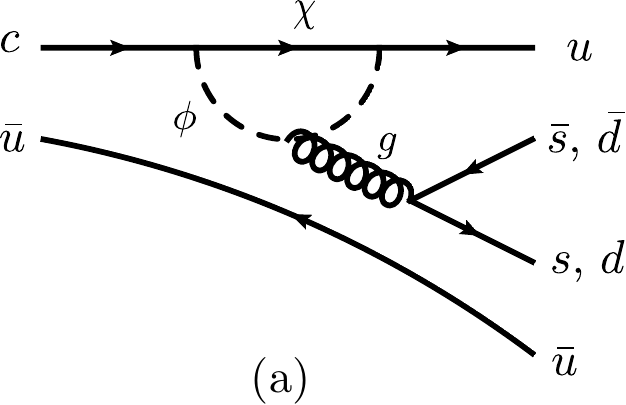}} ~~~~~
\raisebox{20pt}{\includegraphics[width=0.25\textwidth]{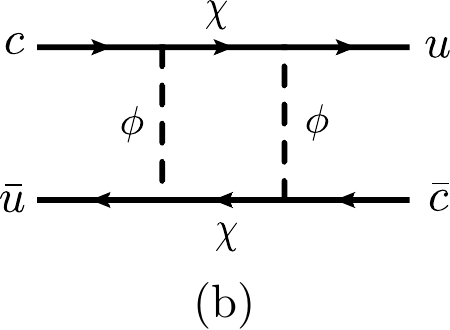}} ~~~~~
\raisebox{29pt}{\includegraphics[width=0.25\textwidth]{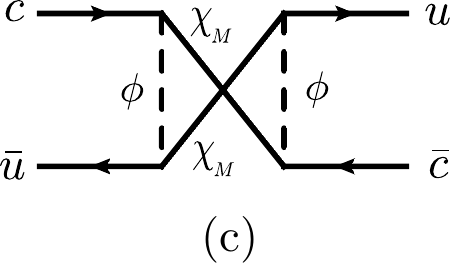}}
\caption{Example Feynman diagrams that contribute to~(a) the $D \to
  K^+K^-$ and $D \to \pi^+\pi^-$ decay amplitudes and~(b and c) $D^0 -
  \bar D^0$ mixing in the discussed models with fermion + scalar
  loops.}
\label{fig:Diagrams_loop}
\end{figure*}

\subsection{Fermion + Scalar Loop without GIM Mechanism} 
\label{subsec:loop_no_GIM}
%
We add one heavy fermion $\chi$ and one heavy scalar $\phi$ to the SM
that couple with right-handed up type quarks
\begin{equation} \label{eqn:noGIM}
\mathcal{L}_{\text{int}} = X_u \bar u_R \chi \phi + X_c \bar c_R \chi
\phi + X_t \bar t_R \chi \phi ~ + \text{ h.c.} ~.
\end{equation}
The couplings $X_i$ necessarily violate flavor and are free complex
parameters. In the following we consider the case where the new
particles are $SU(2)_L$ singlets. We assume the fermion is
electrically neutral and a $SU(3)$ singlet and the scalar is a $SU(3)$
triplet.  Different quantum number assignments do not lead to
qualitatively different results concerning the $D \to K^+ K^-$ and $D
\to \pi^+ \pi^-$ decay amplitudes. We do not consider new fields with
flavor breaking couplings to left handed quarks to avoid constraints
from kaon mixing and $\epsilon^\prime/\epsilon$.

The Lagrangian considered in this framework incorporates an accidental
$Z_2$ symmetry where $\chi$ and $\phi$ are odd under the symmetry
while the SM fields are even. In the considered minimal framework, the
lightest of $\chi$ and $\phi$ is therefore stable and becomes the dark
matter. If the colored scalar is lighter than the fermion, additional
interactions have to be introduced to prevent the scalar from being
absolutely stable. If these additional interactions are weak enough
such that the scalar remains stable on detector scales, bounds from
searches for long-lived particles apply~\cite{Aad:2011mb,
CMS-EXO-11-022}. If instead the scalar decays promptly, then it is
highly model dependent whether or not the considered scenario is
excluded by current collider searches.

In the case where the neutral fermion is the lighter particle, SUSY
searches with jets + $\slashed{E}_T$ at the Tevatron~\cite{:2007ww,
Aaltonen:2008rv} and LHC~\cite{ATLAS-CONF-2011-155, Chatrchyan:2011zy}
provide constraints to the model. In particular, both ATLAS
~\cite{ATLAS-CONF-2011-155} and CMS~\cite{Chatrchyan:2011zy} give
bounds on the production cross section for a simplified model that
contains the first and second squark generations and a neutralino
lightest supersymmetry particle (LSP). We simulate the production
cross section using \verb|Prospino 2.1|~\cite{Beenakker:1996ed} and
set all the superpartner masses to be 4.5 TeV except the relevant
squarks and the LSP, following~\cite{ATLAS-CONF-2011-155}. The
obtained cross section is then scaled by $1/8$ since in our minimal
scenario there is only one colored scalar instead of eight. We find
that neither the ATLAS nor the CMS searches currently put bounds on
our model.

The discussion of the collider constraints is the same regardless of
whether the neutral fermion is a Majorana or a Dirac particle. On the
other hand, the nature of the fermion plays a crucial role in the low
energy phenomenology of the considered model.

\subsubsection{Dirac Fermion}
%
The interactions specified in~\eqnref{noGIM} lead to 1-loop gluon
penguin contributions to the Wilson coefficients of the $\Delta F = 1$
effective Hamiltonian that are shown in diagram~(a)
of~\figref{Diagrams_loop} and lead to
\begin{eqnarray}
\tilde C_6^{(1)} &=& \frac{\alpha_s}{4\pi} X_u X_c^*
\frac{1}{8 M_\phi^2} p(z) ~, \nonumber \\
\tilde C_3^{(1)} &=& \tilde C_5^{(1)} = - \frac{1}{N_c} \tilde C_4^{(1)} = -
\frac{1}{N_c} \tilde C_6^{(1)} ~, \nonumber \\
\tilde C_{8g}^{(1)} &=& X_u X_c^* \frac{1}{4 M_\phi^2} g(z) ~.
\end{eqnarray}
The loop functions depend on the ratio of the fermion and scalar
masses $z = M^2_\chi / M^2_\phi$. We find $p(1) = -1/24$, $g(1) =
1/48$. Their full analytical expressions can be found
in~\appref{loop_functions}.  The relation between the Wilson
coefficients $C_{3,4,5,6}^{(1)}$ of the QCD penguin operators is
universal for all models where they are induced by gluon penguins.

The couplings in~\eqnref{noGIM} also lead, in principle, to 1-loop box
contributions to the $\Delta F = 1$ effective Hamiltonian. These
contributions, however, are strongly suppressed by an additional
factor of $|X_u|^2$ and are therefore negligible.

Finally, 1-loop box contributions to $D^0 - \bar D^0$ mixing are also
induced (see diagram~(b) of~\figref{Diagrams_loop})
\begin{equation}
\tilde{C}_1^{(2) D} = \frac{(X_u X_c^*)^2}{16\pi^2}
\frac{1}{M_\phi^2} \frac{1}{8} f(z)~,
\end{equation}
with $f(1) = -1/3$. The analytical expression for $f$ is given
in~\appref{loop_functions}.

\begin{figure*}[tbp]
\centering
\includegraphics[width=0.45\textwidth]{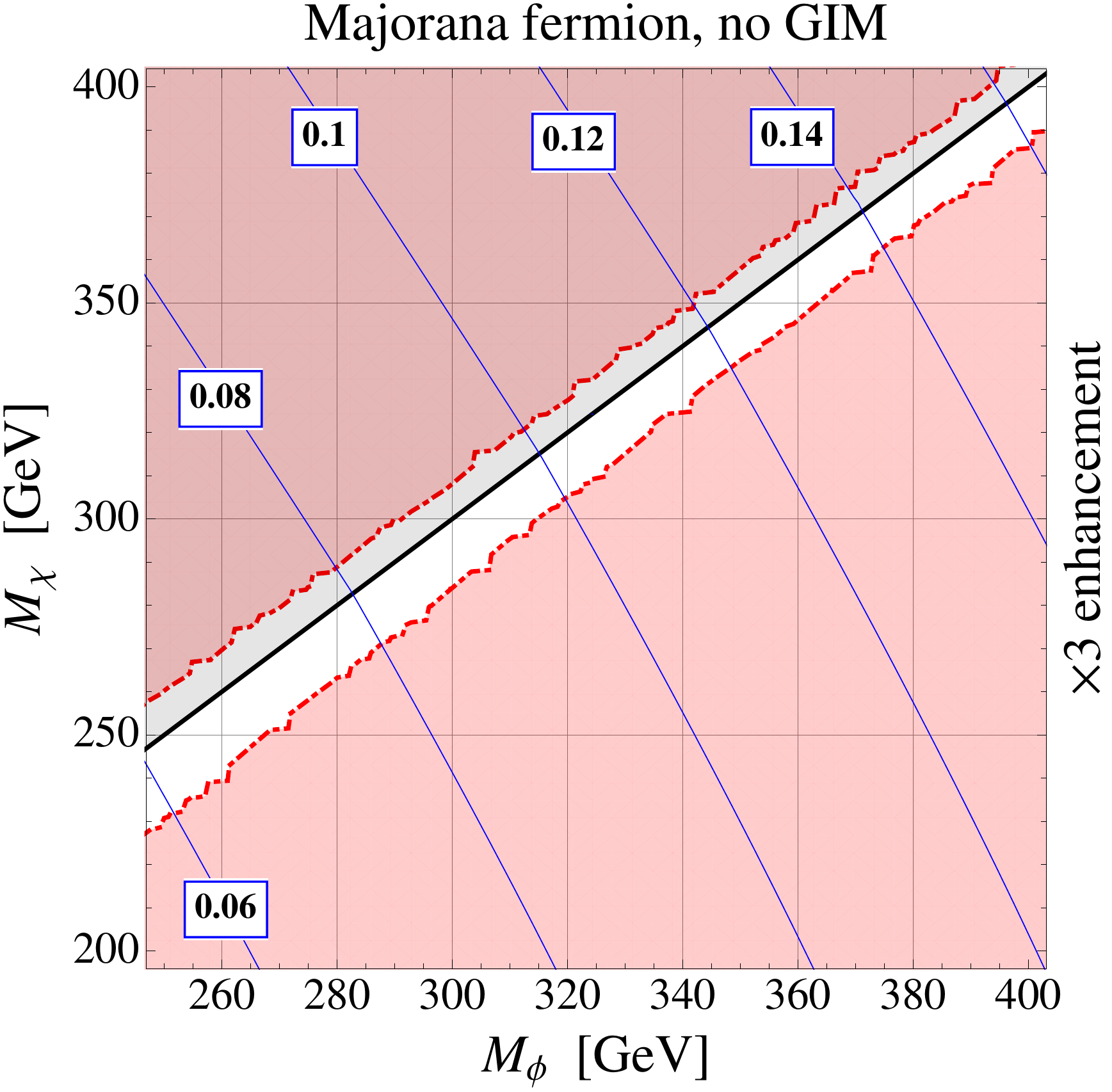} ~~~~~
\includegraphics[width=0.45\textwidth]{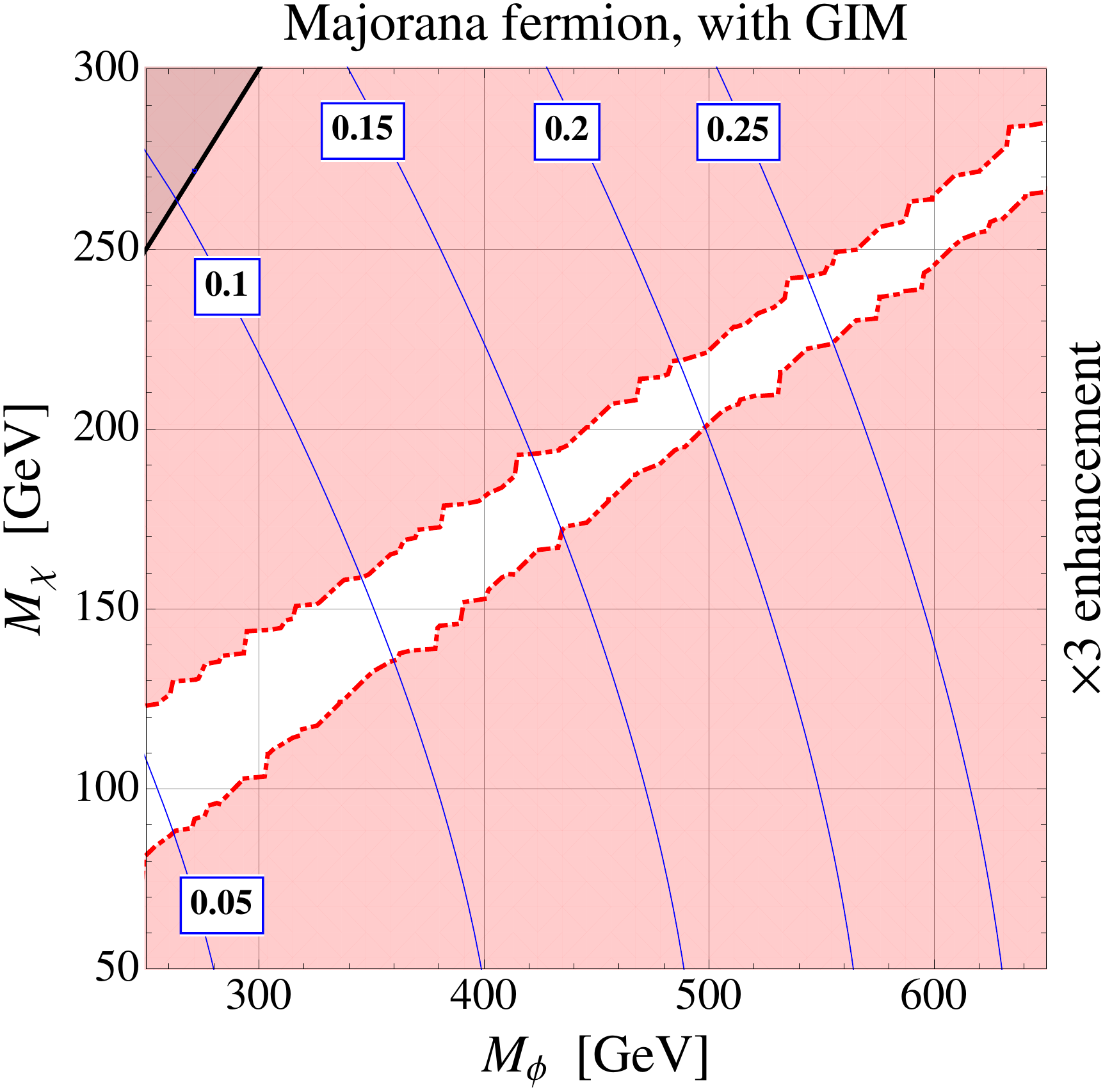}
\caption{The $M_\phi$ - $M_\chi$ plane in the models where a gluon
  penguin is induced by a Majorana fermion - scalar loop without GIM
  mechanism (left) and with GIM mechanism (right). The blue contours
  indicate the (left) $|X_u X_c^*| = 0.06, \ldots 0.14$ or (right)
  $|\delta_{cu}| = 0.05, \ldots 0.25$ values in agreement with the
  measured $\Delta A_\text{CP, WA}$, assuming maximal phases of $\pi/2$. An
  enhancement of the hadronic matrix elements by a factor of 3 is
  assumed in both plots. The red (dash-dotted) region is excluded by
  the $D^0 - \bar D^0$ mixing constraints. In the region above the
  black (solid diagonal) line, the colored scalar is lighter than the
  fermion.}
\label{fig:loop_majorana}
\end{figure*}

We find that $D^0 - \bar D^0$ mixing constraints exclude NP
contributions to $\Delta A_{\rm CP}$ as large as the world average 
even if we allow for an enhancement factor of 3 in the hadronic
matrix elements.

\subsubsection{Majorana Fermion} \label{subsec:majorana}
%
We find qualitatively different results for the Majorana fermion.
While the gluon penguin contributions to the decay amplitudes are
identical for Majorana and Dirac fermions, the box contributions to
$D^0 - \bar D^0$ mixing differ. In the case of a Majorana fermion,
``crossed'' box diagrams also exist (see diagram~(c)
of~\figref{Diagrams_loop}). One finds
\begin{equation}
\tilde{C}_1^{(2) D} = \frac{(X_u X_c^*)^2}{16\pi^2} \frac{1}{M_\phi^2}
\left( \frac{1}{8} f(z) + \frac{1}{4} \tilde f(z) \right)~,
\end{equation}
with $\tilde f(1) = 1/6$. The analytical expression for $\tilde f$ can
be found in~\appref{loop_functions}.

The left plot in~\figref{loop_majorana} shows the $M_\phi$ -- $M_\chi$
plane in the considered scenario. The blue contours labeled with
$0.06, \ldots 0.14$ show the $|X_u X_c^*|$ values that are required to
be in agreement with the measured $\Delta A_\text{CP, WA}$ at the $1\sigma$
level, assuming Arg$(X_u X_c^*) = \pi/2$. An enhancement of the hadronic
matrix elements by a factor of 3 is assumed. The red (dash-dotted)
region is excluded by the $D^0 - \bar D^0$ mixing constraints. In the
region above the black (solid diagonal) line shaded in dark gray, the
colored scalar is lighter than the fermion and would be a stable
colored particle: for this region of parameter space, the model must
be extended to be viable.

We observe that if the Majorana and scalar masses are equal, $M_\phi =
M_\chi$, the constraint from $D^0 - \bar D^0$ mixing can be
avoided. Indeed, for such a ratio of masses, the box and crossed box
contributions to the mixing amplitude cancel. The exact mass ratio
where such a cancellation occurs depends on the quantum numbers of the
scalar and the Majorana fermion. For example, in the well know SUSY
case of squark -- gluino boxes, the cancellation occurs for $M_{\tilde
g} \simeq 1.6 \times M_{\tilde q}$ (see~\cite{Crivellin:2010ys} for a
recent discussion.).  Because of this cancellation, there exists a
narrow region of parameter space where the NP contribution to $\Delta
A_{\rm CP}$ can explain the measured value. The same is in principle
true without assuming any enhancement of the hadronic matrix
elements. In such a case, however the $D^0 - \bar D^0$ mixing
constraints hardly allow any deviation from the $M_\phi = M_\chi$
line.

\subsection{Fermion + Scalar Loop with GIM Mechanism} 
\label{subsec:loop_GIM}
%
We consider now a framework that allows the incorporation of a
GIM-like mechanism. In addition to the heavy fermion, we introduce 3
scalars, $\phi_{u,c,t}$, which are scalar partners of the 3
right-handed up-type quarks.  If the scalars couple universally to
quarks and are approximately degenerate in mass, summing over the
three scalar loops strongly suppresses flavor changing neutral current
processes.  As discussed in~\cite{Giudice:2008uk} for this setup,
$\Delta F = 2$ processes are more strongly GIM suppressed with respect
to $\Delta F =1$ processes and we expect more room for NP in the decay
amplitudes.

We consider the following flavor universal couplings between the
quarks and scalars
\begin{equation} \label{eqn:GIM}
\mathcal{L}_{\text{int}} = g_\phi \bar u_R \chi \phi_u 
+ g_\phi \bar c_R \chi \phi_c + g_\phi \bar t_R \chi \phi_t 
~+ \text{ h.c.} ~.
\end{equation}
In the basis where the quark masses are diagonal, we write the mass
matrix for the scalars as
\begin{equation} \label{eqn:delta}
\hat M_\phi^2 = M_\phi^2 \mathbb{I} + M_\phi^2 \delta~,
\end{equation}
where $\delta_{ij} \ll 1$ are a source of flavor violation. Such a
setup is, for example, realized in the MSSM by gluinos and
right-handed up squarks.  Here, we focus on the case where the new
particles are $SU(2)_L$ singlets, the scalars are $SU(3)$ triplets,
and the fermion is a $SU(3)$ singlet. Different quantum number
assignments do not lead to qualitatively different results concerning
the $D \to K^+ K^-$ and $D \to \pi^+ \pi^-$ decay amplitudes.

Similar to the scenario without the GIM mechanism discussed above, the
framework considered here is only mildly constrained by collider
searches. The difference with respect to the case without GIM
mechanism is that now there are three scalars, and the production
cross section for scalar pair production is therefore three times
larger than in~\subsecref{loop_no_GIM}. On the other hand, one of the
scalars decays predominantly into top quarks and the constraint from
squark pair production with direct decays into light quarks and the
lightest neutralino from~\cite{ATLAS-CONF-2011-155, Chatrchyan:2011zy}
cannot be applied for this scalar. Analogously to the previous
scenario without GIM, we scale the production cross section obtained
by \verb|Prospino 2.1| by the ratio of the number of scalars in the
model, \emph{i.e.} a $2/8$ scaling. We find that the ATLAS
search~\cite{ATLAS-CONF-2011-155} does not put any bound on our
model. The CMS search~\cite{Chatrchyan:2011zy} on the other hand
excludes a small corner of parameter space with scalar masses between
300~GeV and 350~GeV and fermion masses below $\simeq$ 100~GeV.  This
holds regardless of whether the fermion is Majorana or Dirac. The
flavor phenomenology is, however, qualitatively different for the two
cases.

\subsubsection{Dirac Fermion}
%
The 1-loop gluon penguin contributions to the decay amplitudes are
\begin{eqnarray}
\tilde C_6^{(1)} &=& \frac{\alpha_s}{4\pi} \delta_{uc}
\frac{g_\phi^2}{8 M_\phi^2} P(z) ~, \nonumber \\
\tilde C_3^{(1)} &=& \tilde C_5^{(1)} = - \frac{1}{N_c} \tilde
C_4^{(1)} = - \frac{1}{N_c} \tilde C_6^{(1)} ~, \nonumber \\
\tilde C_{8g}^{(1)} &=& \delta_{uc} \frac{g_\phi^2}{4 M_\phi^2} G(z)
~,
\end{eqnarray}
where $\delta_{uc} = \delta_{cu}^*$ is a complex dimensionless parameter
of flavor violation as defined in~\eqnref{delta}.
For the loop functions, we find $P(1) = 1/30$, $G(1) = -1/80$. Their
analytical expressions can be found in the appendix.

Now, 1-loop box contributions to the $\Delta F = 1$ effective
Hamiltonian are not suppressed by additional small mixing angles. For
the minimal set of couplings defined in~\eqnref{GIM}, we find
\begin{equation}
\tilde C_3^{(1)} = \frac{1}{2} \tilde C_9^{(1)} =
\frac{g_\phi^4}{16\pi^2} \frac{\delta_{uc}}{12 M_\phi^2} B(z) ~,
\end{equation}
where $B(1) = 1/48$. Its analytical expression can be found
in~\appref{loop_functions}. We remark that in na\"{i}ve factorization,
this combination of Wilson coefficients does not contribute directly
to the $D \to K^+ K^-$ and $D \to \pi^+ \pi^-$ decays that have either
strange or down quarks in the final state.\footnote{They do
contribute in QCD factorization via annihilation diagrams that we do
not consider here.}  Through renormalization group running, however,
the other QCD penguin operators (in particular $\tilde O_6^{(1)}$) are
induced and lead to non-zero contributions. Although such
contributions are subleading, they are included in our numerical
analysis.

Finally, the 1-loop box contributions to $D^0 - \bar D^0$ mixing are
given by
\begin{equation}
\tilde{C}_1^{(2) D} = \frac{\delta_{uc}^2}{16\pi^2}
\frac{g_\phi^4}{M_\phi^2} \frac{1}{8} F(z)~,
\end{equation}
with $F(1) = -1/30$. The analytical expression for $F$ can be found
in~\appref{loop_functions}.

Similar to the situation without a GIM mechanism, we find that the
gluon penguins that are induced by a Dirac fermion loop cannot viably
account for the measured $\Delta A_\text{CP, WA}$, even if an enhancement
of the hadronic matrix elements by a factor of 3 is assumed.

\subsubsection{Majorana Fermion}
%
The gluon penguin contributions to the decay amplitudes, that are
induced by a Majorana fermion -- scalar loop, are identical to the
Dirac case, but the box contributions to the decay amplitudes and to
$D^0 - \bar D^0$ mixing differ. Adding the ``crossed'' boxes, we find
\begin{equation}
\tilde C_3^{(1)} = \frac{1}{2} \tilde C_9^{(1)} =
\frac{g_\phi^4}{16\pi^2} \frac{\delta_{uc}}{12 M_\phi^2} \left( B(z) -
\frac{1}{2} \tilde B(z) \right)~,
\end{equation}
\begin{equation}
\tilde{C}_1^{(2) D} = \frac{\delta_{uc}^2}{16\pi^2}
\frac{g_\phi^4}{M_\phi^2} \left( \frac{1}{8} F(z) + \frac{1}{4} \tilde
F(z) \right)~.
\end{equation}
with $\tilde B(1) = -1/12$, $\tilde F(1) = 1/20$. Their analytical
expressions can be found in~\appref{loop_functions}.

The corresponding situation in the $M_\phi$ -- $M_\chi$ plane is shown
in the right plot in \figref{loop_majorana}, assuming an enhancement
of the hadronic matrix elements by a factor of 3. The blue contours
labeled with $0.05, \ldots 0.25$ show the $|\delta_{uc}|$ that are
required for agreement with the measured $\Delta A_\text{CP, WA}$ at the
$1\sigma$ level, setting Arg$(\delta_{uc}) = \pi/2$.  Now, the ratio
of the scalar to Majorana mass where the box contributions to meson
mixing cancel is $M_\phi \simeq 2.3 \times M_\chi$. We observe
that constraints from $D^0 - \bar D^0$ mixing leave more regions of
parameter space open, as expected.  Still, the measured value for
$\Delta A_{\rm CP}$ can be explained by NP contributions only in a
narrow region along the $M_\phi \simeq 2.3 \times M_\chi$ line.

\subsection{Chirally Enhanced Magnetic Penguins} \label{subsec:chiral}

\begin{figure*}[tbp]
\centering
\includegraphics[width=0.9\textwidth]{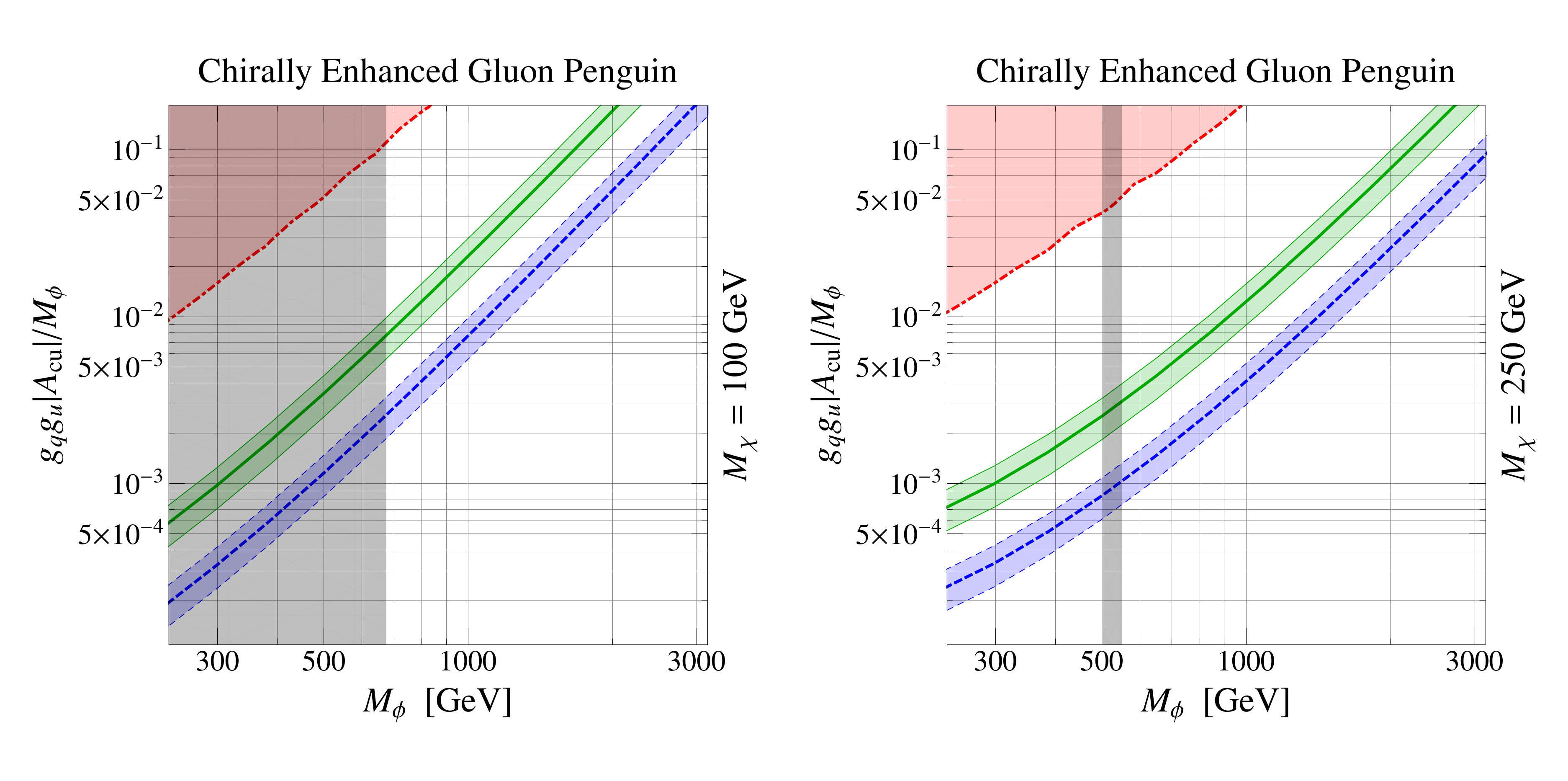}
\caption{Regions in the $M_\phi$ -- $g_u g_q |A_{cu}|/M_\phi$ plane
  compatible with the data for $\Delta A_\text{CP, WA}$ at the $1\sigma$
  level in the model with chirally enhanced gluon penguins, setting
  Arg$(A_{cu}) = \pi/3$. In the left (right) plot $M_\chi =
  100~(250)$~GeV. The green (solid) band corresponds to the expression
  for the decay amplitude in na\"{i}ve factorization, the blue
  (dashed) band assumes an enhancement by a factor of 3. The red
  (dash-dotted) region in the upper left corners is excluded by the
  $D^0 - \bar D^0$ mixing constraint. The dark gray vertical region is
  excluded by jets + $\slashed{E}_T$ searches at LHC.}
\label{fig:loop_chiral}
\end{figure*}

Finally, we discuss a setup that leads to chirally enhanced
chromomagnetic penguins.  We introduce a Majorana fermion $\chi$ that
is singlet under the SM gauge group, as well as scalar partners,
$\phi_q$ and $\phi_u$, to the left-handed quark doublets $Q$ and the
right-handed up-type quarks $U$, respectively. We consider the
following interactions among those degrees of freedom
\begin{eqnarray}
\mathcal{L}_{\text{int, mass}} &=& X_u \bar U \chi \phi_u + 
X_q \bar Q \chi \phi_q + A \phi_u^* \phi_q H  + ~
\text{ h.c.} \nonumber \\
&& + m_u^2 |\phi_u|^2 + m_q^2 |\phi_q|^2 + M_\chi \chi \chi ~.
\end{eqnarray}
We set the coupling matrices $X_u$ and $X_q$ to be universal in flavor
space: $X_q = g_q \mathbb{I}$, $X_u = g_u \mathbb{I}$. We further assume the
mass matrices for the scalars $m_u^2$ and $m_q^2$ are universal and,
for simplicity, also equal: $m_q^2 = m_u^2 = M^2_\phi \mathbb{I}$. The only
new source of flavor violation is then the trilinear coupling
$A$. After electroweak symmetry breaking, the trilinear coupling $A$
leads to mixing between the $\phi_u$ and the isospin $+1/2$ component
of $\phi_q$.  This setup resembles to a large extent the MSSM with
flavor changing trilinear couplings in the up-squark sector as
discussed in~\cite{Grossman:2006jg,Giudice:2012qq}.

Both the elements $A_{cu}$ and $A_{uc}$ can lead to $c \to u$
transitions. Expanding the contribution to the Wilson coefficients in
$A_{cu} v /M_\phi^2 \ll 1$, we find
\begin{eqnarray}
C_{8g}^{(1)} &=& \frac{g_q g_u}{M^2_\phi} \frac{v}{m_c} 
\frac{A_{cu} M_\chi}{4 M_\phi^2} \tilde G(z) ~, \nonumber \\
\tilde C_{8g}^{(1)} &=& \frac{g_q g_u}{M^2_\phi} \frac{v}{m_c} 
\frac{A_{uc}^* M_\chi}{4 M_\phi^2} \tilde G(z) ~, 
\end{eqnarray}
where $z = M_\chi^2 / M_\phi^2$ and $\tilde G(1) = -1/24$. The full
analytical expression for the loop function $\tilde G$ can be found
in~\appref{loop_functions}. We highlight the expected chiral
enhancement of the magnetic penguins by a factor of $v/m_c$.

Box diagrams with the new fermion and the scalars also lead to
contributions to $D^0 - \bar D^0$ mixing. Considering only the
$A_{cu}$ coupling, we find
\begin{equation}
\tilde C_2^{(2) D} = \frac{g_q^2 g_u^2}{16\pi^2} \frac{1}{M_\phi^2} 
\frac{(A_{cu} v)^2}{M_\phi^4} \frac{1}{2} \tilde F(z) ~, 
\end{equation}
where the loop function $\tilde F$ is the same as
in~\subsecref{loop_GIM}. The $A_{uc}$ coupling leads to the analogous
contribution to the coefficient $C_2^{(2) D}$. If both couplings are
present simultaneously,  contributions to $C_4^{(2) D}$ and
$C_5^{(2) D}$ are also generated.

In~\figref{loop_chiral}, we show regions in the $M_\phi$ --
$g_ug_q|A_{cu}|/M_\phi$ plane compatible with the data on $\Delta
A_\text{CP, WA}$ at the $1\sigma$ level, considering $A_{cu}$ as the only
source of flavor violation and setting Arg$(A_{cu}) = \pi/3$. In the
plot on the left (right) we set $M_\chi = 100~(250)$~GeV. The green
(solid) band corresponds to the expression for the decay amplitude in
na\"{i}ve factorization. The blue (dashed) band assumes an enhancement
of the hadronic matrix elements by a factor of 3. The red (dash
dotted) region is excluded by the $D^0 - \bar D^0$ mixing constraints.
The gray shaded regions are excluded by SUSY searches with jets +
$\slashed{E}_T$. Results from ATLAS~\cite{ATLAS-CONF-2011-155} and
CMS~\cite{Chatrchyan:2011zy} indicate that scalar masses up to 675 GeV
are excluded for a fermion mass of 100 GeV (see left plot). On the
other hand, for a fermion mass of 250 GeV, only a small region around
500 GeV is excluded by CMS data alone. For heavier fermion masses, the full
range of scalar masses is allowed by present collider constraints.

Because of the chiral enhancement factor $v/m_c$ in the decay
amplitudes, $D^0 - \bar D^0$ mixing constraints allow for a large
$\Delta A_{\rm CP}$ in the considered setup. Even for scalar masses of
1~TeV and larger, NP contributions to $D^0 - \bar D^0$ mixing are more
than an order of magnitude below the experimental constraints for NP
contributions to $\Delta A_{\rm CP}$ that agree with the experimental
value.

\section{Conclusions} \label{sec:conclusions}

The LHCb measurement of the difference in the time dependent CP
asymmetries in the singly Cabibbo suppressed $D\to K^+ K^-$ and $D \to
\pi^+\pi^-$ decays, $\Delta A_{\rm CP}$, is the first evidence for
charm CP violation. Although there are large uncertainties in the SM
prediction, the measurement could indicate New Physics, and NP
interpretations are nevertheless motivated and exciting.  In this
paper, we studied the effect of NP degrees of freedom for nonstandard
direct CPV in the $D\to K^+ K^-$ and $D \to \pi^+\pi^-$ decays while
also considering constraints both from low and high energy
experiments.

As is shown in~\cite{Isidori:2011qw}, models that give rise to
chirally enhanced chromomagnetic $c\to u$ penguin operators are the
least constrained by low energy data and can easily accommodate the
large $\Delta A_{\rm CP}$ value measured by LHCb. The most prominent
examples for such models are supersymmetric scenarios as discussed
in~\cite{Grossman:2006jg} and very recently in~\cite{Giudice:2012qq}.
We studied the chirally enhanced chromomagnetic penguins in the
framework of a simplified model that contains scalar partners of the
left- and right-handed up-type quarks as well as a Majorana
fermion. We confirm that low energy observables, in particular $D^0 -
\bar D^0$ mixing, as well as collider searches do not significantly
constrain the model's parameter space that leads to a sizable $\Delta
A_{\rm CP}$.

Models that contribute to the $D$ meson decays through four fermion
operators are generically expected to be strongly constrained by $D^0
- \bar D^0$ mixing data~\cite{Isidori:2011qw}. Nonstandard effects in
the decays are only possible if the new degrees of freedom mediating
the $c \to u$ transition are very light. In this work, we quantified
this statement through a systematic study of models with a minimal set
of new degrees of freedom giving rise to four fermion operators both
at the tree and the loop level.  In summary, we find:

\begin{itemize}
\item Flavor changing couplings of the the SM $Z$ boson can induce a
$\Delta A_{\rm CP}$ as large as the observed value if the NP phase is
moderately tuned to avoid constraints from indirect CP violation in
$D^0 - \bar D^0$ mixing.
\item A $Z^\prime$ that mediates the $c \to u$ transition at tree
level cannot account for the observed $\Delta A_{\rm CP}$ due to the
combined constraints from $D^0 - \bar D^0$ mixing and dijet searches.
\item A heavy gluon with a flavor changing tree level $c \to u$
coupling and with a mass of 200~GeV $\lesssim M_{G^\prime} \lesssim$
320~GeV cannot fully be excluded as NP explanation for the measured
$\Delta A_{\rm CP}$ if a moderate enhancement of the hadronic matrix
elements is allowed.
\item In a 2HDM with MFV, there exist regions of parameter space that
can lead to a sizable $\Delta A_{\rm CP}$. They are characterized by
light charged Higgs masses and strongly enhanced couplings of the
right handed strange quark to the charged Higgs, with respect to its
SM Yukawa coupling. Avoiding constraints from perturbativity and $B
\to X_s \gamma$, however, requires a considerable amount of fine
tuning.
\item Scalar octets can also induce large nonstandard effects in a
mass window 200~GeV $\lesssim M_{\phi_8} \lesssim$ 320~GeV that is
left open by current collider searches. The viable parameter space is
analogous to the 2HDM with MFV model and appears to be, to some
extent, fine tuned.
\item The scalar diquark model we consider is ruled out by $D^0 - \bar
D^0$ mixing as a NP explanation of $\Delta A_{\rm CP}$.
\item The minimal models that induce NP effects in the $D$ meson
decays through loops with Dirac fermions and scalars are strongly
constrained by $D^0 - \bar D^0$ mixing data and cannot give rise to a
sizable $\Delta A_{\rm CP}$.
\item If Majorana fermions and scalars appear in the loops, then $D^0
  - \bar D^0$ mixing constraints can be avoided for a particular ratio
  of Majorana and scalar masses that depends on the exact quantum
  number assignment for the particles. Correspondingly, in such models
  there exist regions of parameter space that lead to large $\Delta
  A_{\rm CP}$ in agreement with the data.
\end{itemize}

We note that our results are robust, since changes to the central
value of $\Delta A_{\rm CP}$, new direct search constraints, and the
enhancement or suppression of flavor bounds coming from additional
field content can be readily applied to our minimal models and our
derived Wilson coefficients.

As we showed, the New Physics parameter space favored for an
explanation of the LHCb evidence for charm CP violation is largely
within the current reach of various direct searches at the LHC.  Thus
the well-known complementarity between low energy flavor measurements
and high energy direct probes may prove fruitful as we continue to
search for New Physics.  Our work emphasizes this synergy by
presenting a broad study of minimal New Physics models, discussing
both their effects on low energy flavor observables as well as their
high energy collider signatures.  We have demonstrated that a number
of intriguing New Physics models can viably explain the large $\Delta
A_{\rm CP}$ measurement, and we have concretely isolated the
interesting parameter spaces of such models which must now be searched
directly.

\bigskip
\paragraph*{Acknowledgments:}
%
Fermilab is operated by Fermi Research Alliance, LLC under Contract
No. De-AC02-07CH11359 with the United States Department of Energy.
R.P. thanks the National Science Foundation for support under Grants
PHY-0757481 and PHY-1068008.  R.P. and C.-T.Y are supported by the
Fermilab Fellowship in Theoretical Physics.  The authors would like to
thank T.~J.~Khoo for useful information regarding the ATLAS
jets+$\slashed{E}_T$ search, Andy Cohen and Martin Schmaltz for useful
comments, and Daniele Alves, Jonathan Arnold, Bogdan Dobrescu,
Stefania Gori and David Straub for useful discussions.

\appendix
\section{Hadronic Matrix Elements in 
\texorpdfstring{Na\"{i}ve}{Naive} Factorization} 
\label{sec:HME}
%
To evaluate the hadronic matrix elements of the operators in the
$\Delta F =1$ effective Hamiltonian in~\eqnref{Heff_DF1}, we use
na\"{i}ve factorization
\begin{eqnarray}
&& \langle K^+ K^- |(\bar u \Gamma_1 s)(\bar s \Gamma_2 c)| D^0
  \rangle \nonumber \\
&& \simeq \langle K^+|(\bar u \Gamma_1 s)| 0 \rangle \langle K^-|(\bar
  s \Gamma_2 c)| D^0 \rangle ~,
\end{eqnarray}
where $\Gamma_i$ represent the various Dirac and color structures. In
this approximation, it is straightforward to evaluate the hadronic
matrix elements $\langle O_i \rangle \equiv \langle K^+ K^- |O_i| D^0
\rangle$
\begin{eqnarray}
\langle O_1^{(1)} \rangle &=& N_c \langle O_2^{(1)} \rangle = N_c 
\langle O_3^{(1)} \rangle = \langle O_4^{(1)} \rangle \\
&=& -2N_c \langle O_9^{(1)} \rangle = -2\langle O_{10}^{(1)} \rangle 
\nonumber ~, \\
\frac{1}{\chi_f} \langle O_1^{(1)} \rangle &=& N_c 
\langle O_5^{(1)} \rangle =
\langle O_6^{(1)} \rangle = -2N_c \langle O_7^{(1)} \rangle 
\nonumber \\
&=& -2 \langle O_8^{(1)} \rangle = -8\langle O_{S1}^{(1)} \rangle = 
-8 N_c\langle O_{S2}^{(1)} \rangle ~,\nonumber \\
\langle O_{T1}^{(1)} \rangle &=& \langle O_{T2}^{(1)} \rangle = 0 ~, \nonumber
\end{eqnarray}
where $N_c = 3$ is the number of colors and $\chi_f$ is the
appropriate chiral factor from \eqnref{chiralfactor}.  Using QCD
factorization methods, the matrix element of the chromomagnetic
operator is~\cite{Grossman:2006jg}
\begin{equation}
\langle O_{8g}^{(1)} \rangle = - \frac{\alpha_s}{4\pi} \frac{N_c^2 -
  1}{N_c^2} (3 + \chi) \langle O_1^{(1)} \rangle ~.
\end{equation}

As QCD conserves parity, the matrix elements of the chirality flipped
operators $\tilde O_i^{(1)}$ are identical to the ones shown above.

\section{Anomalous Dimensions} \label{sec:gamma}
%
For completeness we collect here all the anomalous dimensions of the
$\Delta F = 1$ operators that are required for our analysis. The LO
anomalous dimension matrix that governs the running and mixing of the
current-current operators $O_{1,2}^{(1) p}$, the QCD penguin operators
$O_{3,\ldots,6}^{(1)}$ and the QED penguin operators
$O_{7,\ldots,10}^{(1)}$ is given by (see {\it
e.g.}~\cite{Buchalla:1995vs})
\begin{widetext}
\begin{equation}
\gamma^0_{1,...,10} = \begin{pmatrix}  
\frac{-6}{N_c} & 6 & \frac{-2}{3N_c} & \frac{2}{3} & \frac{-2}{3N_c} &
\frac{2}{3} & 0 & 0 & 0 & 0 \\
6 & \frac{-6}{N_c} & 0 & 0 & 0 & 0 & 0 & 0 & 0 & 0 \\
0 & 0 & \frac{-22}{3N_c} & \frac{22}{3} & \frac{-4}{3N_c} &
\frac{4}{3} & 0 & 0 & 0 & 0 \\
0 & 0 & 6 - \frac{2f}{3N_c} & \frac{-6}{N_c}+\frac{2f}{3} &
\frac{-2f}{3N_c} & \frac{2f}{3} & 0 & 0 & 0 & 0 \\
0 & 0 & 0 & 0 & \frac{6}{N_c} & -6 & 0 & 0 & 0 & 0 \\
0 & 0 & \frac{-2f}{3N_c} & \frac{2f}{3} & \frac{-2f}{3N_c} &
\frac{6(1-N_c^2)}{N_c} + \frac{2f}{3} & 0 & 0 & 0 & 0 \\
0 & 0 & 0 & 0 & 0 & 0 & \frac{6}{N_c} & -6 & 0 & 0 \\
0 & 0 & \frac{d-2u}{3N_c} & \frac{2u-d}{3} & \frac{d-2u}{3N_c} &
\frac{2u-d}{3} & 0 & \frac{6(1-N_c^2)}{N_c} & 0 & 0 \\
0 & 0 & \frac{2}{3N_c} & -\frac{2}{3} & \frac{2}{3N_c} & -\frac{2}{3}
& 0 & 0 & \frac{-6}{N_c} & 6 \\
0 & 0 & \frac{d-2u}{3N_c} & \frac{2u-d}{3} & \frac{d-2u}{3N_c} &
\frac{2u-d}{3} & 0 & 0 & 6 & \frac{-6}{N_c}
\end{pmatrix} ~,
\end{equation}
\end{widetext}
where $N_c =3$ is the number of colors, $f$ is the number of active
quark flavors, and $u$ and $d$ are the numbers of active up- and
down-type quarks, respectively.

Leading order running of the chromomagnetic operator $O_{8g}^{(1)}$ is
given by
\begin{equation}
\gamma^0_{8g} = \frac{4 N_c^2 -8}{N_c} ~.
\end{equation}
For the LO anomalous dimension matrix responsible for the running and
mixing of the scalar and tensor operators $O_{S1,S2}^{(1)}$ and
$O_{T1,T2}^{(1)}$, we find
\begin{equation}
\gamma^0_{ST} = \begin{pmatrix} \frac{6-6N_c^2}{N_c} & 0 & 
\frac{1}{N_c} & -1 \\
-6 & \frac{6}{N_c} & -\frac{1}{2} & \frac{2-N_c^2}{2N_c} \\
\frac{48}{N_c} & - 48 & \frac{2N_c^2 -2}{N_c} & 0 \\
-24 & \frac{48-24N_c^2}{N_c} & 6 & \frac{4N_c^2+2}{-N_c}\end{pmatrix} ~,
\end{equation}
which agrees with~\cite{Buras:2000if} once the different conventions
for the operators and the $\sigma_{\mu\nu}$ matrix are taken into
account.  At leading order and in the limit of massless down and
strange quarks, the scalar and tensor operators do not mix into other
operators.

As QCD conserves parity, the anomalous dimensions for the chirality
flipped operators $\tilde O_i^{(1)}$ are identical to the ones
shown above.

\section{Loop Functions} \label{sec:loop_functions}

The loop functions $h_1$ and $h_2$ appear in the charged Higgs
contributions to kaon mixing in the 2HDM in~\subsecref{2HDM}
\begin{eqnarray}
h_1(x) &=& \frac{1+x}{2(1-x)^2} + \frac{x}{(1-x)^3} \log(x) ~,
\nonumber \\[10pt] 
h_2(x,y) &=& \frac{x-4y}{(1-x)(y-x)} + \frac{3y^2
  \log(y)}{(1-y)(x-y)^2}\nonumber \\
&& + \frac{2xy-4y^2+x^2(3y-1)}{(1-x)^2(y-x)^2} \log(x) ~. \nonumber
\end{eqnarray}

The loop functions $p$, $g$, $f$ and $\tilde f$ appear in the
discussion of the model with fermion -- scalar loops without a GIM
mechanism in~\subsecref{loop_no_GIM}. The functions $p$ and $g$ occur
in the expressions for the gluon penguin contributions to the $D$
meson decay amplitudes. The function $f$ comes from the evaluation of
a box diagram contributing to $D^0 - \bar D^0$ mixing and $\tilde f$
comes from the corresponding crossed box diagram
\begin{eqnarray}
p(z) &=& -\frac{2-7z+11z^2}{36(1-z)^3} - \frac{z^3}{6(1-z)^4} \log(z)
~, \nonumber \\
g(z) &=& \frac{1-5z-2z^2}{24(1-z)^3} - \frac{z^2}{4(1-z)^4} \log(z) ~,
\nonumber \\
f(z) &=& -\frac{1+z}{(1-z)^2} - \frac{2z}{(1-z)^3} \log(z) ~,
\nonumber \\ \nonumber \\
\tilde f(z) &=& -\frac{2z}{(1-z)^2} - \frac{z(1+z)}{(1-z)^3} \log(z)
~.
\end{eqnarray}

The loop functions $P$, $G$, $F$ and $\tilde F$ are the analogues to
$p$, $g$, $f$ and $\tilde f$ in the fermion -- scalar model with a GIM
mechanism discussed in~\subsecref{loop_GIM}. The functions $B$ and
$\tilde B$ appear in the box and crossed box contributions to the $D$
meson decay amplitudes of that framework:
\begin{eqnarray}
P(z) &=& \frac{1-5z+13z^2+3z^3}{18(1-z)^4} + \frac{2z^3}{3(1-z)^5}
\log(z) ~, \nonumber \\
G(z) &=& \frac{-1+8z+17z^2}{24(1-z)^4} + \frac{z^2(3+z)}{4(1-z)^5}
\log(z) ~, \nonumber \\[10pt]
B(z) &=& \frac{1+5z}{8(1-z)^3} + \frac{z(2+z)}{4(1-z)^4} 
\log(z) ~, \nonumber \\
\tilde B(z) &=& \frac{z(5+z)}{2(1-z)^3} + \frac{z(1+2z)}{(1-z)^4}
\log(z) ~, \nonumber \\[10pt]
F(z) &=& -\frac{1+10z+z^2}{3(1-z)^4} - \frac{2z(1+z)}{(1-z)^5} 
\log(z) ~, \nonumber \\
\tilde F(z) &=& -\frac{z(17+8z-z^2)}{6(1-z)^4} -
\frac{z(1+3z)}{(1-z)^5} \log(z) ~. \nonumber
\end{eqnarray}

Finally, the loop function $\tilde G$ appears in the expression for
the chromomagnetic penguin loop in~\subsecref{chiral}
\begin{eqnarray}
\tilde G(z) &=& -\frac{1+5z}{4(1-z)^3} - \frac{z(2+z)}{2(1-z)^4} 
\log(z) ~. \nonumber
\end{eqnarray}



\end{document}